SSRC ITIC 2001 Work Paper

## E-Governance, International Cooperation and Security - New Millennium Challenges for a Small Country.

Neki Frasheri
SSRC ITIC Program Fellow - Summer 2001

### 1. Introduction

Talking about Information and Communication Technologies (ICT) and their impact on developing countries (DCs), about the information society, electronic-governance and potential "leapfrogs" of these countries into a prosperous future, all this is becoming a common thing for researches, journalists, politicians, and sociologists. Perhaps it is talked more about the possible futures and less about the present, and how this present may condition the future. One thing is sure - information and communication technology is changing the world, enriching and integrating communication means worldwide breaking all geographical and social borders. It is important to consider ICT as a key tool in the hands of humanity, and not as an actor playing a key role in the history. It is creating conditions for more economical and political freedom, which may lead to new movements and institutions for democracy [Joshi 1999]. We are at the beginning of a new technological revolution whose consequences is difficult to evaluate. The ICT impacts get shaped as result of the fusion of globalization, worldwide connectivity and knowledge networking [Choucri, 2000]. A revolution makes new power structures to arise over old ones; and always technology has been the catalyst, not the cause [Wriston, 1997b].

In the paper we make a criticism of different views on the supposed role of ICT for the future of human society. This criticism is seen from the point of view of a small developing post-communist country as Albania, hoping that the conclusions would throw some light for developing countries in general, especially those in a transition stage. We examine some aspects of international collaboration and security, where the ICT implemented in the public administration may have an important impact. Understanding the role of public administration and the structure of its interfacing with the public and NGOs is the next subject, to be followed by the discussion of e-governance issues. Based on these arguments, the development policies and practices are examined, including relations between public and private sectors. Following arguments of many authors, we identify or redefine some crucial factors that negatively impact the role of ICT in the development of the country, its relations with the international community, and ways to push forward its development.

Of course DCs have big differences with western developed countries. But even between DCs themselves there are big differences, due to the geographical position, history and cultural heritage. Albania is a developing country situated in a good geographical position. For a long time it remained isolated from the rest of the world in a kind of total "communist self-colonial regime" (as defined by the journal Nature during the 1970s). During the last ten years it is undergoing a turbulent transition time switching from dictatorship to anarchy. In this paper we try to analyze the challenges the ICT in the public administration (public administration) and the e-governance present for countries as Albania, with particular attention to international collaboration and security.

The past and recent ICT challenges of Albania are described elsewhere (Beqiraj and Frasheri 1998, Frasheri 2000). Nevertheless, it is useful to say a few words as background. Staying in isolation for many years, Albania was forced to make almost everything by itself, and so it was developed a considerable industry, agriculture and, what is more important, enormous human resources in many fields as geology, engineering, computer science, agriculture, health care, chemistry. The development was strongly polarized due to political and ideological factors. This polarization was between developed technology and human technical knowledge in one side, and rigid planning and management in the other side. The control mechanism of society was the political feedback, compared with the market feedback in western countries [Salomon et al., 1994].

Technical capabilities could not make way before rigid policies based on an imaginary ideological reality. This situation created an enormous gap between the Albanian society and the rest of the world, which borders were just "a throw of a stone" away. The total isolation was so strong that political leadership of that time was able neither to make any change nor to control the situation. Even the reverse of communism was done in a scandalous way, followed by some massive bursts of emigration towards neighbor countries.

The country entered in a transition stage towards the democracy and free market economy in anarchist way. The industry and agriculture didn't fail from the free concurrence, but were destroyed under the name of the "shock therapy". The "shock therapy" as applied in Poland is analyzed by Comisso [1999], who describes it as "*the immediate and simultaneous introduction of liberalization and stabilization in a single package of policy measures*." But Comisso adds that this strategy in itself was debatable, together with the fact that the results of Polish economy were due to this strategy or to its modifications done subsequently. The presence of such debate reflects the controversy of the "shock therapy", and implies that priority must have the "therapy", that is a good package of rules and procedures to open the way of enterprises for independent activities in a free market but refraining from anarchy. In Albania they did only the "shock" of the economy destroying it, by opening the market in one side and blocking the activities of enterprises on the other. When regulations do not create conditions for a democratic development of the economy, a government monopoly may become a private monopoly [Braga, 1998]. On the other hand, Colomer [1995] considers Albania as a case where communists were able to control democratic reforms. Transition was transformed in a complex of negative phenomena. The public administration was made chaotic as result of liberalization, combined with massive movements of population from remote regions toward big cities. Flourishing research, development and engineering work was reduced drastically. Many people left or are trying to leave the country, between them many good specialists, creating a critical hole within the society. All this happens in a time when politics pretends for development of the country and its integration within international frameworks.

In the field of telecommunications, Albania inherited an old and out of date network. But in the field of IT a good basis was built; beginning with the first introduction of computers in 1971, continuing with the creation of the Institute of Informatics and Applied Mathematics and of the first metropolitan network in 1985, and with the Department of Informatics at the University of Tirana. A characteristic of this development was that IT was used mainly for engineering tasks. While the management and the economy, conditioned very strongly by politics, very little used it except some statistical data processing. During the transition the engineering work was interrupted or reduced to the use of PCs, so nothing sustainable remained in place from the past. The transition period is characterized by "dispersed" investments and development. The telecommunication network has been rebuilt in main cities, and new optical links are installed with neighbor countries. Now in big cities is possible to use dial-up connections for Internet and data communication. The mobile telephony is in extension to cover the whole country, and the second operator is expected to start within the year 2001. As result of an improvement of the marketing policy by introducing prepaid cards, mobile telephony is becoming quite popular. Some ICT is implemented in central public administration institutions; new projects are in place for its improvement, as well as for the extension in local administrations. Education on IT is widespread, and even included experimentally in the program of high school. The country is experiencing a real technological leapfrog…

The concept of "leapfrogging" is widely analyzed by authors as Davison, Vogel and Harris [2000]. Technological leapfrogging is considered as a direct deployment of new modern technologies, without using step-by-step previous technologies already out of use in developed countries. It is already accepted that ICT are something as "inevitable luxury" for DCs. Even when not necessary, a minimal up-to-date infrastructure in DCs is obligatory for sake of compatibility with developed countries. Moreover, today it may be more difficult to find, use or maintain old technologies. There were many cases in Albania when foreign organizations made as donation old infrastructures they had to throw away. In all the cases these infrastructures were not used or used very little; it was simply because the equipment was old, nobody knew how to use it, something was missing, or it was incompatible with existing new equipment. The key question is not about the technological leap, but in what impact creates such technological leapfrog the development of society. Bangemann Report [Bangemann et al.1994] presents the actual development as "*throughout the world, ICT are generating a new industrial revolution… this revolution adds huge new capacities to human intelligence and… changes the way we work together and the way we live together*." It means that "leapfrogging" for DCs must be considered as something more than simply technological advance. In our conditions, until the technological leapfrog "changes the way we work together and the way we live together", there is no leap towards developed countries of the present, and the information society of future. Looking around we see that the impact of new ICT is negligible, the country is dominated by non-constructive politics, the regional situation is politically and military quite hot.

This makes us to think more deeply about the ICT in public administration, about e-governance, and its impacts on collaboration and security issues. The question is not simple - will the country go forward or not. In some aspects it may be forced to go forward. But moving forward while widening the gap of digital-divide within the country will have grave consequences for the Albanian society in general, and for the regional collaboration and security as well. In a polarized society, the poor layers of population serve as background for all kind of illegal activities and conflicts, which are know well in our region in these recent years. In a developing society, there are many new political forces that emerge from "nothing", and the recent history of Balkans is a good example how such new policy may throw in fire entire regions. When polarization and developing come together, that is the prelude of explosion. The history seems to consider (and not only in Balkans) the thesis of Huntington, supposing civil clashes with the emergence of a new order [Vital 1999]. Such extremist phenomena may happen not only in DCs, even in quasi-unified Europe certain political forces claim for at least federalism.

During the year 1999 a survey was done in the framework of the Y2K program in Albania. In all government institutions (except the directorates of taxes and of customs), the situation resulted that there was no IST system in place that would create problems regarding the change of the date. This "good" situation was result of lack of institutional IS implemented over the existing IT, the latter used mainly for individual data processing. One year after a new survey was carried out, aiming to evaluate how [in-house] IST people in public administration think about their own situation. Due to the lack of conceptual knowledge that prevails in public administration, and comparing with the results of the Y2K survey, one may conclude that these results represent an over-estimation of the reality. Some results of survey are presented in this paper.

## 2. "Towards Information Society" - where are we going?

To think about ICT used in public administration and related issues concerning international cooperation and security, it is necessary to evaluate the reality where public administration "lives", and the perspectives of this reality. Already there is an enormous quantity of writings by many scholars about the role of ICT transforming the human society towards a future "information society". Many supposed attributes of this future society are analyzed, considering the ICT as a universal tool for democratization. The opinions vary from one extreme to the other, and all this reminds of a folk story when the wise man of village said "or it is something, or it is going somewhere", when saw a big bug crawling on the grass. Joshi [1999], for example, considers the information society as organized "horizontally" based on voluntarism, while the actual society is organized "vertically" in hierarchical classes. But at the same time Joshi admits that ICT revolution is itself a contradictory phenomenon that "*breaks down hierarchies and creates new power structures*," empowering both the control and the anarchy of society.

Making possible a fast and free worldwide exchange of information, ICT serve as a tool of acquiring knowledge, a non-material kind of wealth. While the Industrial Revolution emphasized the physical mass, Information Revolution has the opposite effect emphasizing a non-physical mass as information [Mathews, 2000]. Wriston [1997b] points out the new emerging competition for information, mentioning rich countries like Singapore and Hong Kong that have virtually no physical assets. This phenomenon theoretically implies the development of humanity in a "knowledge society". Nath [2000], for example, theorizes on the ways such society will be governed, suggesting that this society will be more free and have more opportunities to decide from whom it will be governed and how. At the other extreme stays Guehenno [Kaase, 2000], who considered the ICT revolution the end of democracy. While Bimber [1998] has a critical position; he argues that changing of informational environment in wired societies is not a sufficient condition to change the interest of people in public affairs. Bimber proposes instead the idea of "accelerated pluralism". In [Bimber 2001], he argues that a positive correlation between development of ICT and political participation of citizens is not supported by historical trends in US; and that ICT perhaps affects political participation through cognitive phenomena more than through increase of information flows.

Wrighton [1997b] describes the impact of ICT as increasing the power of individuals and outmoding old hierarchies in all components of the order that emerged from the Industrial Revolution (national sovereignty, economy, and military power). But, continues Wriston, "*despite all of the advantages of science and the ways in which it is changing the world, science does not remake the human mind or alter the power of the human spirit.*"

Actually there are many embryonic examples how different initiatives and projects lead people to learn more about their governments and improving their reaction for a better government. Referring at the site http://www.digitalgovernance.org/ in June 2001, such cases include: e-procurement for business opportunities (Chili), free Internet access for citizens through tele-centers (Costa Rica), on-line voting (Estonia), comparing electoral programs site (France), land property registration (India), issue of civil status and real estate papers via cyber-kiosks (India), on-line audits for villages' accounts (India), video-conferencing facilities for high rank officials (India), services for payments, licensing, certification (India), dissemination of information (India, South Korea), sites for litigation cases of high courts (India), keeping trace of environment performances of ministers (India), controlling custom/tax revenues (India), increasing public awareness against corruption (Kenya, Philippine, UK, US), smart identity cards (Malaysia), general information (Pakistan), customs on-line (Philippine), dissemination of legislative documents (South Africa, Vietnam), facilities for people to fax their local members of parliament (UK), petitions on-line (US), development of democracy (Bangladesh, Burma, Costa Rica, Ecuador, Ghana, Zimbabwe), and other cases of international organizations. Such simple cases e-governance may be found in many other countries, including Albania as described in following sections. Evaluating these cases, we may conclude that:

- First, examples of dialogue between citizens and governments are mainly in developed countries (UK, US), where the development of the society, governance and democracy are consolidated gradually in centuries. Or it is in "half-developed" countries (India, Malaysia, South Africa) where the average of development of the society may be low, but there are some communities highly developed, and/or there is a consolidated economical class that controls the politics and administration. In both cases there is the public administration that, willing or not willing, becomes interested to promote those activities. Such conditions may not be valid for many DCs, due to the lack of will in public administration, or popular masses are not so strong to push administration to accept it, or other problems may have priority.

- Second, in many cases ICT is used simply as an alternate media for dissemination of information - "printed" CDs instead of printed paper, or some centers where people may go and consult a database instead of asking an official for information. This is a good step forward, but is far from transformation and democratization of governance, as we will see below. For example, printing of bulletins from public administration may be costly (when compared with "printing" of CDs); but the paper has a certain legal value and it can be used even in conditions of lack of ICT, quite usual in DCs. While to give legal value to web sites and databases, it is problematic both technically (security and maintenance of servers) and legally (new legislation, introduction of e-signature etc.).

How good would it be the future information society, or we have simply some optimistic scenarios? As we will see below, application of ICT may make governance more transparent in some aspects, as well as it increases means of propaganda, ways of spying and terrorism etc. [Vital 1999]. Transparency itself is only a necessary condition for democracy - even communist governments knew how to be transparent in certain conditions; and their [partial] transparency resulted part of "brain-washing" processes. Choucri [2000] remarks the actual trend of two components of democracy, i.e. the demand site (citizen) and the supply site (government), which is not converging. The reality seems to be more complex for "early optimism" [Eichengreen and Kohl 1998, Kaase 2000]. If we stick on what we know, taking into account that "mathematically" extrapolating in the future increases the margins of error exponentially, we may say that:

i)    Widespread of ICT applications is changing radically our world, our work and our living in community.
ii)   ICT is creating a worldwide public space, breaking all borders of space and time. This public space is being used increasingly for all communication-related activities of humanity.
iii)  All communication-related problems of humanity are extending in this new public space that seems to be without borders. More important it will be as public space; more [known] human problems will emerge there.

Many authors, pointing out many cracks in "optimistic" scenarios, discuss the "optimistic/pessimistic" question. Heeks [2001b] writes that much emphasis is given for usage of ICT in business (e-business) and NGOs areas; while other socially important areas of e-governance, related with poor countries and poor communities are neglected giving way for transforming the digital divide in a information and knowledge divide. Referring the phrase "information society" thrown by Daniel Bell in 1973 as significance of ICT for emergent service-dominated economies in post-industrial societies, the Margetts [1991] concludes that: "*The principle behind the idea of an information polity is one of transformation through integration of government and IT*". The question is not if there will be a transformation or not, but in which direction this transformation may be and what impact it will have for different countries and communities.

Authors of [NRC-CSTB 1997] go deeper in analyzing of possible scenarios. The ICT are becoming available for lower layers of societies as result of "low cost information and communication resources". But, continue the authors, beside the availability low cost ICT, not all trends lead to decentralization of information sources. Electronic mass-media ownership has tendency to become more concentrated, favored by deregulation principles of actual governing ways. This concentration leads to creation of very big operators ("CNN phenomenon"). Such very big operators may influence deeply in policy-making, and their weight may overshadow possibilities of new ICT used by individuals or small communities. Latzer defined even the concept of "mediametics", as the fusion of telecommunication, computing and electronic media [Kaase, 2000]. Chester [1998], after remembering the story of Orsen Welles and its broadcast of "War of Worlds" in 1936, concludes that today, due to the development of ICT, "*the phrase 'over to our reporter in terrain' may now mean a reporter in the next room with a virtual image of the terrain in background*".  All these arguments tend to prove that deployment of ICT remains a contradictory process that oscillates between distribution and concentration of power.

We need to take into account specifics of broadcasting versus browsing ICT. Broadcasting activities (i.e. TV) are "active services", people simply select the channel and acquires what is broadcasted. People may have many channels to select, but in any case the multitude of channels is linked with important hardware installations that favor big operators. While Internet offers a "passive service" that does not requires important investments, where people need to "browse actively" to get the information. This gives freedom of selection, but at the same time creates conditions for miss-orientation of attention in socially non-relevant issues, depending on individual hobbies. In the age of satellites, electronic mass-media may remain the most active service for dissemination of information, and combined with "mega-sites" of Internet may lead in new ways of e-government, which would be not so democratic as expected. Without active attitude of people in social and political issues, the future may be a new kind of dictatorship, and this reminds us about the importance of human resources.

We have already many examples of Internet control for political purposes by governments (cases of China, Singapore), of contradictory political attitudes towards the multimedia networks as a tool of development and democracy (case of Malaysia, [Joshi 1999]). The answer for the future is not very simple - the history will make its own way in disfavor of some government, but new realities may lead to new civil clashes and creation of new "governments".

The principle "the big one has a stronger voice" becomes relevant with the increasing of the Internet, for the simple reason that people depends more and more on search engines as the only independent way of finding the sites of interest in "cyberspace". Choucri [2000] remarks that without search engines the Internet would be worse that a "plate of spaghetti", it is matter of space, time and quality. In [NATO 1999], authors argue that the time of information gathering in the digital age is a critical factor for decision-making and international collaboration and security. Authors remark that search engines are able to solve somehow the problem, but not sufficiently. They identify two problems with search engines. The one is that they scan the cyberspace indiscriminately, which gives enormous quantity of indexed pages. The other is that search engines do a selection based on the level of usage of pages. The web-space covered by search engines is evaluated in 30-60% of the whole; there is no distinction on quality and reliability of the information; and with the increase of data volumes, the time of indexing becomes relevant and certain documents may loose their importance when included at least into searching indexes, as well as important documents may be "lost" because of not used.

On the same issue of search engines, the idea of neutrality of search engines is itself debatable. While search engines remain a powerful tool of accessibility in the Web, but many scholars [Golding, 1994; Pollack, 1995; Schiller, 1995; Hoffman and Novak, 1998], point out that due to social, economic and racial factors, the Web is pre-configured in political ways that result in exclusion of many voices from being visible through a cyberspace distorted, due to *biased search engines*, in favor of circles wealthy in economical and technical resources (see [Introna and Nissenbaum 2000]). The politics of search engines reflects the traditional political struggle in a new "digital environment", to sustain the democratic character of traditional media and Internet as well. Lawrence and Giles (1999) have estimated that individual search engines studied by them index not more than 16% of the total Web (estimated in 800 million index-able pages) [Introna and Nissenbaum 2000]; and through the combination of several search engines, the coverage of Web increases about 42% (that is the average of estimation claimed by [NATO 1999]). This fact is already pointed out by end-users who find the Web almost inconceivably large so even search engines are only partially effective on finding interested data. The question is not for what we may find using search engines, but for what we many not, and the latter is related with the position of marginal communities and developing countries - due to the lack of financial and technical means, these communities even present in cyberspace, may remain pretty invisible.

The contradiction between search engines covering only a part of the cyberspace, and the remaining invisible "dark matter" is only a part of a greater one between centralization and decentralization trends of development.

This leads us to two crucial questions - how it will be the integration between IT and government, and what would be the consequences of this integration. Human society development has always gone through a differentiation process, even under the logo "*liberte, egalite, fraternite*".  We have already some bad recent examples regarding theoretical scenarios and realities. Communist regimes were presented as a leap towards better societies; what happened in practice was a kind of "political capitalism" that replaced the old financial capitalism, and the results were a global disaster for the countries that experimented this "good" idea. Would "information society" be some new kind of "information capitalism"? Many authors are already discussing the phenomena of "digital divide", and it is a very important issue of consequences the information revolution may have. We may distinguish two cases of "digital divide":

i)   Real digital divide, that is result of being not online, out of cyberspace, as it is the case of some marginal poor communities.
ii)  Virtual digital divide, that is result of being online but invisible in cyberspace, as is the case of some communities with little economic and technical resources.

Both cases apply in DCs, and the second case may apply even in developed countries. Heeks [2001a], for example, argues that "*E-governance lies at the heart of two global shifts: the information revolution and the governance revolution*". This process changes the ways how the society works and is governed, creating good opportunities for effectiveness and good governance, but all this is privilege of being online and having access to digital information and knowledge related with governance reforms. Heeks concludes that the e-governance gap is "*increasingly separating developed and developing countries, and elite and ordinary citizens within developing countries*".

Another author, Chester [1998], remember us that "*none of the previous shifts in society were made without social consequences*". Chester continues that the key problem of information society is connected with the so-called "ideology of consumerism", that is the intensive use of ICT to gain production efficiency using information as capital. This ideology leaves no place for other social values, and technology serves in "*maintaining a culture of unrestrained capitalism rather than a new society*". Going into more concrete issues, Curry [2001], for example, suggests that maybe the Internet will open up weak economies to competition from Web sites in more advanced nations with far fewer reciprocal opportunities DCs, having as consequence shifts and surprises with unforeseen results. The issue is analyzed by many scholars, and Garcia [2001] concludes on the necessity for marginal communities to "*reengineer themselves to meet the requirements of a knowledge-based network economy … [they] must integrate their economic activities, and thicken their institutions by reinforcing their local and regional ties*."

The problem complicates also due to barriers of language, literacy, localism characteristic for many DCs and communities within, which create gaps between formal and informal knowledge systems [Gupta A., Kothari B., Patel K. 2000]. The debate for multilingual domain names structure of Internet is a good example of "virtual digital divide" due to barriers of language. It creates the possibility for some countries and communities to be more Internet-active, but at the same time it extends in cyberspace the language barrier between themselves and other countries. Automatically they become invisible for the rest of the world. Internet serves to connect people worldwide, when certain language and cultural conditions necessary for communication are fulfilled. But the cultural and linguistic barriers in cyberspace may not be the worst case of being marginalized; the worse would be for certain categories of people within the country, who are already marginalized due to differences in age, culture, education, poverty.

The issue of remaining invisible is quite concrete for marginal communities and DCs. As we will see further, there is a strong trend to "go private", that is "consumerism", and that this trend is supported even by international organizations, which often are a powerful driving force in DCs [Heeks 1998a]. In this context, the question "where we are going?" implies not hypothetical scenarios, but real risks for the society, especially in DCs. It is why Chester [1998] argues that to avoid a revolution (based on consumerism) and better address the interests of the society as a whole, it is necessary a well planned and managed evolution process, taking account of resulting social consequences. This conclusion moves us from a "technological terrain" to a "political terrain". Cohen, Delong and Zysman [2001] observe that cyberspace will be the place of policy-making in the future, regarding fundamental organization issues of society and of market. *"Under the best circumstances the policy risks are high"*, conclude the authors.

The background of risks is related with the complexity of the situation. Castells [1999] argues that ICT is changing our world, but it is not ICT the cause of these changes. Historically the humanity depends on the information, and new technologies, only change the means and ways of its circulation. In this context, the e-governance is result of adoption of new ICT by governments. This adoption implies profound changes on information resources and information flows, changing inter-government and citizen-politician-state relationships [Bellamy and Taylor 1994]. All these arguments point out the decisive role of governments in e-governance processes. Hamelink [1999] says that the challenge for both governments and international community is sustainable development through use of ICT conditioned not by technological factors but by political decisions. Beside crisis and ideology elements for a reform, Heeks [1998a] remembers also about the third element - the political will to do the reform. Mathews [2000] concludes that ICT revolution is shaped by wise or stupid policies and social choices.

 A typical positive example is described by Diamond [Diamond 2000] in the case of Bishkek City Hall reform (Kyrgyzstan). It was the political will of the mayor and its supporters that made possible the reform itself and the collaboration with the civil society for this purpose. Scenario of Kosovo compared with that of Bosnia is another example of the decisive character of political will in crises. Skoric [1996] remembers that with the beginning of military activities, all normal communications between Croatia and Serbia were interrupted, and latter the links with Bosnia were destroyed; all this made very difficult the communication between people working in different sides of fighting. While in Kosovo the links were totally interrupted, the only resource of information was through refugees escaping in neighbor countries and air photography.

The role of ICT in international collaboration and security is dependent of the politics - it may change the ways of making politics but not its content. Numerous examples show that Internet is becoming a tool for the organization of activist groups; at the same time governments use ICT to control or monitor this activism. Activist groups use ICT to fight for democracy, human rights and protection of environment, bypassing government channels. Terrorist groups use the same technology to fight for political power against democracy. This dualism has to do with the content that runs over the technology.  The fact that the technology makes governance more transparent worldwide breaking the borders of space and time, does not mean that we new society will be "unified". It is probable that many actual borders may be fall and separate things may be unified. At the same time it is probable that new separation lines may emerge, new "virtual countries" may be created, new political activities emerge replacing the old schemes we are used today. See foe example Mathews [1997b].

Coming back to the Albanian reality, - it is difficult to say how good is the political will. The complexity of events during ten years of transition leaves a bitter taste. The Krantzberg's First Law cites that "technology is neither good or bad, nor it is neutral" [Brodnig and Schonberg 2000]. Deployment of new technologies in social phenomena may be contradictory, by improving something while destroying faster something other, leading the society towards civil clashes [Hill, 1996]. Application of new technologies in DCs may involve less people and have less positive impact in the general development of the country. It may lead to new polarization of society and new social and political crises in the region. Internet as a tool for democracy may become a tool for de-stability.

How governments will react before the "invasion" of ICT, it depends on how they consider their citizens:

"partners", "tools" or "raw material". In a country where, due to economical and social development, government considers its citizens as "partners", we may expect good e-governance and the application of democratic principle "people lets leaders to govern them" [Nath 2000]. Where citizens are considered as "tools", we may hope that something on e-governance would be done to improve the governance including services for citizens. Where political forces consider people as simply "raw material" and do not hesitate to throw into the fire of civil wars their citizens, nothing good may be expected by application of ICT for e-governance - it would be "hell-governance". Even in the past there were means by which leaders would leave people to let them govern. We cannot escape from the "political will". It is just the political will, built over particular economical interests, that controls the development of countries, and it will control the emerging e-governance. In local scale, it is supposed that social consequences would be addressed in the interest of society as a whole [Cester 1998], and managed on four fronts: technological, environmental, human resources and organizational [Willcocs 1994]. All this is totally conditioned by the economic-social-political structure of the country, ICT have nothing to do with it - it is simply a tool to manipulate data.

In global scale the situation is pretty contradictory. Developed countries and international organizations are making many efforts and investments in many DCs, both for development and resolving of local crisis. On the other hand, the global policy is characterized by a blind importation of western rationality. In this context Avgerou [Avgerou 2000] argues that the western rationality may be used to define a range of technical and economical problems in DCs, but it seems quite unsuccessful to force solving of these problems in DCs according to western rationality. IT is considered as one of most important forces for the development of DCs, and Avgerou considers this technology-deterministic point of view as naïve but also neglected by majority of researchers and professionals, missing to "*address deployment of ICT-based IS as manifestation of different rationalities.*" Monroe [2001] on the other side proposes even a shift form the rational choice theory towards the perspective theory, based on the supposition that perception of ourselves in relation with others defines the available options we may choose.

A great danger is the "cargo cult," that may arise where the developing economies observe the benefits ICT bring to the industrialized nations, relying on the blind belief that similar benefits will quickly accrue to themselves, and hurry to acquire the same technology [Davison, Vogel and Harris 2000]. Lessons may be learned from other countries, but there is no "one best way" - each country must be helped to find its "own" way, emphasizes Heeks [2001b]. Conclusions of Avgerou and others show that not always rationalities of DCs are taken into account. Even in technological issues we see the tendency of the different producers, normally from developed countries, to drive end-users towards consumerism. The fast development of ICT seems to be conditioned more by pure concurrence than real user requirements - technology is developed and afterwards it is considered how to use it. This phenomenon is reflected in cases of decreased turnover of ICT producers and failure of Application Service Providers. Western rationality seems to have its own "cracks".

More important to understand is the contradictory character of the technology itself - the same issues that made the Internet to "go there where no network has gone before" [Cerf 1996], make it a good terrain for problematic activities. Monitoring of information flows is more difficult with the multiplication of information routes. TCP/IP as a flexible protocol for transparent reliable exchange of data packets through complicated networks of any topology makes the control of data packet transits more difficult.  The same factors making this technology required by people as a universal communication medium make its control desirable for governments. ICT creates greater possibilities for information access and governments may be forced to accept the availability of these possibilities for the people in order to obtain the advantages of technology. But this does not solve the problem; it only shifts it to another "dimension". People have more information but also miss-information becomes more difficult to be distinguished from reliable information. In certain sectors this phenomenon may create unpredictable complexities (in health care, for example) [NRC-CSTB. 1997]. To assure reliability of services and information, networks may be controlled. It depends on who controls and why.

This control can be used for both motives - to build or to destroy - "technology is neither good or bad, neither is it neutral". The case of Y2K showed clearly the extend and the depth of problems related with increasing use of ICT in every field of human activities. Today it is possible to control remotely the behavior of end-users, by the same means used for protection against hacking (see [Iyengar 2001] for example). If the end-user is permanently on line he can be controlled remotely, perhaps from places where the user country's law cannot be applied; and if he goes through a proxy or a firewall, he is totally dependent by the administrator of the site that can monitor him using simply logging facilities of the system, the same facilities used to identify hackers. Even today many free programs we download from the Internet automatically modify the configuration of computers without being asked to do so, including links to specific sites or installing specific plug-ins. The problem is not simply on privacy issues that someone may use for discrimination in the real word. It is technically possible to differentiate users and give them manipulated information depending on user's personality. Differentiation of users is a reality now, at least in commercial dimensions (see for example [Kopytoff 2001]). Governments are already controlling Internet backbones for criminal activities [Schwartz 2001]. Browsing through the Internet we leave traces that someone can exploit, the most simple consequence is the well known spamming. From a specific point of view spamming, as a "denial of service" attack, represent the best example how the impact of ICT as a mean of acquiring information may be neutralized simply by overloading with information.

Even well appreciated black boxes as routers resulted with glitches that may permit their remote unauthorized control [CERT ADVISORY CA-2001-14]. But commercial producers will not use anyway open software because of confidentiality of their technologies. Computer viruses got a new dimension with the deployment of world wide networking. The case of default "user-friendly" configuration of some email client software, "guilty" for worldwide spread of recent viruses via email, is a simple indication what may happen in the future using commercial software. On the other hand, open software source is accessible by all and a good programmer may modify it to monitor or control end-users. The only partial escape remains an open market where the user may select the hardware and software. But actually there is only one major producer of micro-processors, and only one major producer of operating systems. When wired we enter in a "virtual dictature". If in a political dictature we know who is [formally] at the top, in a "wired virtual one" it may be invisible, and even governments are not immunized against this phenomenon.

What has been for a long time a fiction, it is risky to become reality. It started with "individual hackers" and it is ending with "governmental hackers". In 1984 Brenton and Beneich published a fiction book how a democracy tested cyber-war against a dictature, and how that dictature planned to use cyber-war against its own people. Only four years latter the Internet was shacked by the Robert Morris Jr. "Internet worm". With the deployment of the Internet, numerous cases were reported, when hackers had break in even important governmental sites. In June 2001 news from Reuter [Wolf 2001] cites a report presented in the US Congress about the preparations some countries for cyber-war, considered as a new kind of military operations. And this time "*God is [not] always good for the big battalions*" [Voltaire, see Wriston 1997b]. Even "what is war" and "who is enemy" now seems fuzzy. While information flows are blurring the notion of geographical boundaries, perhaps new "boundaries" may emerge between ideas of "we" and "they" [Mathews, 2000] - that is new virtual territories.

## 3. E-Governance challenge.

Development of e-governance through intense use of IST by public administration, and interacting with citizens and civil society, it represents a solution not only for good and transparent governance and accountability (see [Heeks 2001a]). Interaction between public administration and citizens through networking contributes on putting the governance under critics of the society, improving coordinated actions between government and NGOs, decreasing the possibilities for manipulations of the public and better orientating activist groups. The key for such achievements lies in the information systems of public administration as the core of e-government. The process is pretty contradictory, because dissemination of information is a threat for power structures. ICT changes the way wealth is created and, as consequence, changes ruling elites that control the society [Wriston, 1997a].

In Fig.1 we present ICT-based communication in Albanian public administration, as seen from in-house IST people. It summarizes the exchange of data within and with outside, both in floppies and through networks.

Combined with the results presented in Fig.4, the conclusion is that ICT is little used for exchange of data within organizations, and less with outside. The summary of results would be:

• First, the use of ICT mainly individual.

• Second, IST people do not understand very well the difference between individual use and institutional use. First of all, good governance requires good evaluation of the reality through processing multiple data, to produce the necessary information for supporting decision-making. It is responsibility of organizations, linked with each other in the framework of public administration structure, to exchange and process multiple data. The individual work is only a necessary condition for functioning of different nodes of administrative structures; to get a sufficient condition we need to add the capacity for data exchange between different nodes both horizontally and vertically, as well as their integration in organization scale to produce information.

Heeks [2001a] defines e-governance as "*the use of ICT to support good governance*", improving information exchanges between government and citizens. While automation in the past addressed the internal work of governments, e-governance implies the transformation of external work of governments in their relations with citizens. Following arguments of Heeks, we may theorize about the functionality of ICT within public administration in an e-governance environment as in the Fig.2.

Public administration with its IST has a central role in the framework of e-governance. In the Green Paper [1999] of the European Commission there is pointed out that: "*public sector information plays a fundamental role in the proper functioning of the internal market and the free circulation of goods, services and people.*" Moreover, this paper points out also that: "*without user-friendly and readily available administrative, legislative, financial or other public information, economic actors cannot make fully informed decisions*." The medium layer - INTERFACE - is the sub-system that connects public administration with citizen. This layer and its structure are crucial for communications between administration and citizen.

The conclusion from Fig.2 is that to have real e-governance, IST of public administration in all its levels must guarantee an effective circulation of information with outside - both NGOs and citizens. It implies that the work has to be based on public inter-connected databases implemented or mirrored in the border of public administration layer. Bangemann Report [Bangemann et al. 1994] argues "*interconnection of networks and interoperability of services and applications are recommended as primary Union objectives*."

The network may be used in many ways for exchange of information between individuals or groups; but when the system becomes very complicated such individual dialogues based on e-mail-like services are destined to fail. If we remember the history of Internet, the World Wide Web technology was designed to build distributed databases easily and "universally" usable. Zysman and Weber [2000] point out the fact that primary characteristics of Internet, compared with other proprietary networks as Minitel (France), is that the Internet development is defined and controlled by its users. In our context it is necessary to have public information systems user-oriented and open for web-like interfacing data interchange with their environment, in order to have conditions for good governance and good impact of ICT. The mailing lists or discussion groups must be seen as a complementary component to be used for specific exchanges within limited audiences.

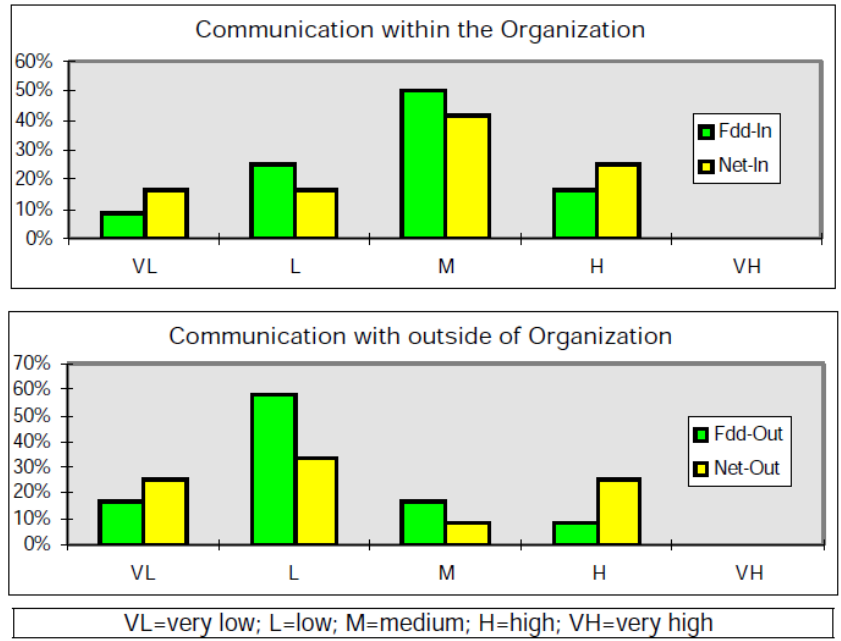

Fig.1 - ICT-based communication in public administration.
(Usage of floppies [Fdd] and networking [Net])

The IST of public administration must create collaboration possibilities with surrounding environment. This cannot be done without decentralization - centralized systems in a distributed environment cannot have any future. The human society in itself is a distributed system, and the failures of ex-socialist "centralized" countries are a proof of "centralized" failures. But development of IST for e-governance in public administration becomes a complex process due to the decentralization and interaction with the outside environment. Chester [1998] warns about the great complexity of links between information, technology and society, the IS itself being in the core of the society. While Landsbergen and Wolken [2001] point out the role of interoperability for the success of ICT deployed in the "*experiment*" of e-governance, as being difficult to be implemented. They argue also that if interoperability is achieved, "*it will result in a fundamentally different way of doing government*". Interoperability is thus a key issue in understanding how information technology will truly affect the public sector.

The recent experience in Albania shows a negative phenomenon - "pseudo-decentralization". In paper local administrations and universities have certain autonomy, in practice they strongly depend on central institutions. This makes the relations' center-periphery complicated and unpredictable. Following the analysis of Ridge [1994], we may conclude that with the decentralization increases the need of interoperability between autonomous IS implemented in different levels of government. It has to do with the balance between "monitoring" and "controlling" - decentralization means improvement of monitoring while releasing direct controlling. To have positive results and avoid unpredictable developments, it is necessary to challenge the difficulties of decentralization through interoperability, and not to solve strategic decentralization problems through operative re-centralization.

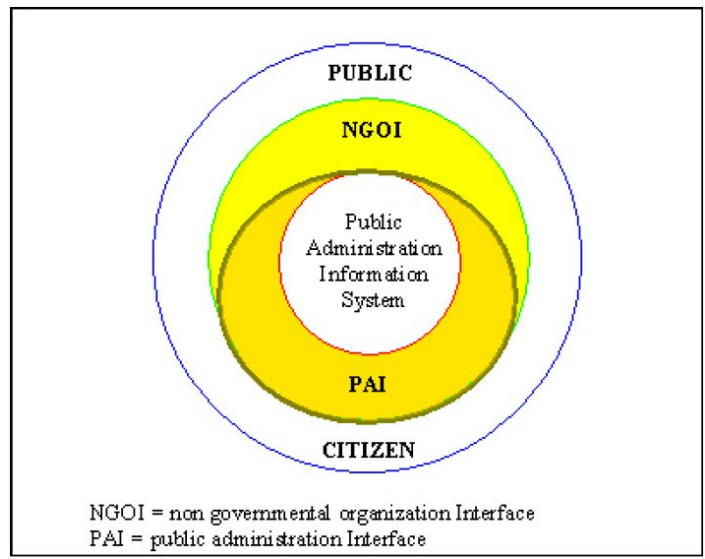

Fig.2. Interfacing public administration with citizens.

Ways of decentralization depend on political objectives. Opening of governance toward citizens, and decentralization affect political structures and relations. Related with this argument, Taylor and Williams [1991] point out that usage of ICT in governance raises important issues about horizontal and vertical organization of public administration. Such issues include vertical organization related with decentralization, and horizontal fragmentation of government into agencies. While Hood [1990] and Willcocs [1994] suggest that fragmentation of management structures within a certain organizational and technological environment may lead to contradictions if separate processes do not converge. This implies that decentralization of the public administration and its open informatization need to harmonize with each other. The practices of giving autonomy with one hand and taking the control with the other only penalize the development, and it has to do only with political issues - a hidden trend to strengthen monopolization of politics in a distributed environment through controlling the interoperability. Concentration of power will lead to a polarization of the society and increasing of risks for civil clashes..

Bardhan and Mookeherje [2000] suggest possibilities decentralization creates for reduction of corruption, particularly in DCs. It is argued that central governments do not have the adequate local information to properly monitor local officials. Myging [1998] argues also about creation of powerful criminal networks as an important element of societies in DCs. Myging emphasizes the effect of such criminal networks on discouraging foreign investors. Heeks [1998b] argues that IT may create new opportunities for corruption at the same time while detection and neutralization of old opportunities. Use of ICT in a distributed environment challenges the status quo of existing "who gets what" [Landsberg and Wolken 2001]. We would like to consider decentralization as a mean for anti-corruption as a positive scenario, because the trend of local authorities of countries in transition may be to get profits from their position as much and as soon as possible. This leads to the creation of certain cupolas that take the control of every significant activity within their territory. Officials of public administration may be attracted towards to extreme poles: monopolization or negligence of responsibilities, and this suffocates the climate of cooperation and development.  In some cases it may lead even to conflicts with the central authorities in the framework of emergence of new political forces. In such cases the consequences may be disastrous for a democratic development. Decentralization is like a knife with two edges for DCs. The balance between decentralization, interoperability and centralization is a vital but difficult issue.

In this context we may conclude that one way to decrease the corruption is well definition of tasks and work procedures for public administration in center and in periphery, and it's opening to the public through the use of ICT. Many authors [including Landberg and Wolken 2001, Heeks 2001a] identify information services as a mean of effectiveness, efficiency and responsiveness for new working ways as: within government permitting joined-up thinking, between government and public by strengthening the accountability, within communities supporting concerted actions and building social and economic development. This argument leads to hope that by forcing the development of distributed IST in public administration, despite its low usage, there are created conditions that may lead to improvement of exchanges between citizens and government, increasing empowerment of citizen and democracy. But it is not a sufficient condition.

The same Heeks [2000] also points out that Internet being a political space, its usage does not lead automatically to empowerment and democracy. It is up to the civil society to democratize the Internet-space through activism, in particular by serving as interface for marginalized communities. This process of democratization relies in the democratization potential of NGOs, both in local and international scale. The role of international NGOs is increasing in the framework of globalization. Madon [2000] concludes that, in the framework of the ideological ascendancy of neo-liberalism and globalization, international NGOs are becoming an important factor influencing the global policies addressing development and democratic issues, serving as a link between local communities and global institutions.

Behind these arguments on NGOs and e-governance lie the status of marginal communities and the role of NGOs to improve the communication with them. NGOs may serve to connect marginal communities with the global network, serving as "physical interface" to overcome infrastructure shortages, and as "logical interfaces" to overcome lack of knowledge and technical education. The problem of marginal communities is crucial for countries as Albania, where poor communities represent a considerable part of the population, and serve as background for all king of illegal activities. To have international collaboration and security, especially in regional scale, it is necessary to integrate these communities in the global development and decrease the polarization of the society. Integration means to talk to them but also to listen from them [they also produce information (see [Heeks 1999a], [McConell 2000]). Actually we see the involvement of NGOs on creation of interfaces of communication between local governments and marginal communities (see following section for details), and this is an important component of the e-governance.

NGOs will impact the progress of DCs. The results of their impact will depend on how NGOs itself will develop, and how they will build relations with the politics. Revolution of ICT has cracked government's monopolies, leading for globalism. At the same time ICT amplifies social and political fragmentation, by enabling more identities and interests to emerge worldwide. In this framework NGOs are widening considerably their fields of interest. Facilitated by global low-cost communication means, NGOs are increasing pressure on governments for many issues, including security and collaboration ones [Mathews, 1997a]. Perhaps it will lead to new ways of policy-making and new shape of political forces. It reminds the old postulate of socialist times, that "organizations of masses are branches of the party"…

## 4.  Public Administration, International collaboration and Security

International collaboration and security (international collaboration and security) are becoming today one of key political issues, which importance deduces from the trends of globalization and the potential of ICT to speed up this globalization. A typical example is the International Relations and Security Network, as an initiative to promote the free flow of unclassified information about security policy issues in the Euro-Atlantic region [Forum 1998]. Several projects are developed under this initiative, aiming on integration of distributed databases and their interfaces, dedicated for international relations and security issues. Governments themselves have contributed in making *holes* through domestic-foreign affairs borders, as result of different international accords and political practices, permitting international interventions in internal problems related with democratic development, human rights protections and international security [Mathews, 1997b]. These issues need to be into consideration when addressing the ICT policies and practices in public administration.

We will try to classify international collaboration and security issues. This classification would help to understand better the role of public administration and ways of its development. We will distinguish two categories of these issues: international collaboration and security in (1) *peacetime*, and in (2) *crisis-time*. The analysis takes into consideration the fact that in all these processes public administration interacts with the public and NGOs in different levels of governance.

The role of public administration and its information systems for international collaboration and security in *peacetime* would include:

- War against criminality and fraud, through international collaboration against illegal activities (related with the police, borders control, customs, taxes),
- Social and health protection, through improving of living conditions of marginalized people, nation-wide health care, promotion of local development through economic activities and exchanges, promotion and exchange of cultures, education etc.,
- Environment protection and improvements, related with pollution, deforestation, water resources, wild life etc., for an optimal exploitation of local and regional assets.
- Preparation, dissemination and application of legislation, regulations procedures and standards, especially those related with daily activities of citizens and cross-border activities.

In this framework it is necessary to include the role of public administration as a leader in undertaking legislative and administrative actions necessary to open the way for the integrity and reliability of IST and their content. Braga [1998] emphasizes also the importance of proper regulations conditioning the ability to benefit from new technologies and attract investments, as well as to promote development even in remote areas. Without access to global communications, citizens and economies will be in difficulties before a global economy. In the future the concept of "being well-informed" would mean not only to have information, but also to have reliable information, and in this context the accountability of service providers and mechanisms for certifying the reliability of information will be crucial [NRC-CSTB. 1997]. In cyberspace it is not easy to answer the crucial question: "who owns what?" That answer has historically been at the roots of all social clashes. The global chaotic use of Internet does not mean democracy - anarchy and democracy are two different things. In DCs the social conscience of ordinary people is less developed, especially in relations with government entities, by strongly neglecting or overestimating it. In a "wired" society, ordinary people will make responsible, for any big fraud related with unreliable information, their government for not keeping things under control; while the same people would have been all the time against the control from the government due to principles of democracy and privacy. In DCs such situations may serve for explosion of civil clashes, as happened in Albania in 1997, when people asked government for the money lost in private pyramidal schemes, and that was the beginning of the riots, anarchy and destroying. (See Braga [2000] for example). Trust and confidence on cyber-systems is an important but undervalued issue [Feldman, 2000].

With the globalization trends of today, we expect the change of public administrations' role, by giving more way to other organizational entities as for example NGOs. The reality will force governments to change their structures and behaviors. From economical (and legal) point of view, today is very difficult to define an asset when consists on information products, because information resources are not bound by geography and not easily controlled by governments. From communication point of view, it is known the case of W.Willson who ordered the total control of communication cables between USA and Europe during the peace conference after the WWI; compared with the case of G.Bush who understood that could not control any more the flow of information across borders, and used the potential of the world information free market during the Gulf War [Wriston, 1997b]. Globalization is changing our world and its governance, and DCs are involved in "their own way" in this process. Strong emigration trends, intense economical exchanges and movement of people in cross-border regions, which countries as Albania are experiencing, are an example of globalization processes. Direct links and cooperation between public administrations of cities form different neighbor countries is another example of cross-border decentralized activity. The developments in one country have impact in other neighbor countries and wider.

In terms of information issues, the globalization followed by other phenomena as democracy and, decentralization creates new problems in security matters, both for developed and developing countries in their own ways. In DCs the legislation and its application are not consolidated as in developed ones, and this makes more difficult the regulations of information flows that may have strong negative impacts in society. Not only cyberspace is "overloaded" with information without guaranties for its quality and reliability, but low costs of electronic publishing allow anyone to put their information and the Internet is becoming playground of all kind of organizations and groups that may provide questionable or misleading information [NATO 1999]. Such phenomena may lead more to miss-understandings rather than reducing it [Bibmer, 2001].

The Information revolution is forcing the acceleration of decision-making processes. In DCs that deploy ICT, without improving old mentalities and social-political rules in the right way, dissemination of misleading information and acceleration of decision-making may lead to unexpected and uncontrollable rapid social and government reactions, and to unpredictable crises. In this context the role of public administration, even legally being limited within a well-defined territory, in reality extends beyond it and takes an international character in all the aspects. It is not simply the war against illegal activities that requires international collaboration. By improving the dissemination of information, risks for non-predicted crises decrease, and cross-border exchanges improve. It has a relaxing effect between different communities and ethnic groups involved in these exchanges, and it reduces the ethnic exploding potential in mixed-up regions as Balkans. The experience of Balkans in these last years showed how delicate are the inter-ethnic relations after a long history of clashes; and how easily conflicts, as those Balkans is actually suffering, may explode.

Arguing on the second category of problems in *crisis-time*, we need to distinguish two sub-categories of crises: natural crises and political crises.

Natural crises are related with different cases of catastrophic events as earthquakes, floods, environment damage, health problems (epidemics) etc. Disaster management in case of natural crises involves many organizations and networking becomes crucial for successful disaster response. Analyzing activities for implementation of ICT (GIS) systems on disaster management in India (state of Maharashtra, after the earthquake of 1993), Vatsa [Vatsa 2000] concludes that IST applied for disaster response provides a new paradigm in connectivity and database availability; and that capabilities to deal with information flows and decision-making significantly improved. Sometime the consequences of natural crises may be difficult to understand and their long-term character may require even years for the community to accumulate the evidence [Landberg and Wolken, 2001], and in such cases the inter-operability between IST of different organizations may facilitate the evidence of problems saving time and lives. The public administration is an active actor, which collaborates with the population and NGOs - local and international, to undertake the measures to respond to natural disasters. This process needs a global coordination of independent organizations, i.e. a "centralized management" into a "distributed environment". The only natural way to achieve it is the intense exchange of information between independent actors.

Political crises are more complicated, and not simply because politics itself complicated. Cross-border information flows erode the difference between domestic and foreign, as well as between government and marketplace. Governments and activist groups can reach directly foreign citizens, and a kind of "trans-national" civil society is emerging. In this emerging information society may have place new ethnic, religious, nationalist, and interest-based backlashes [Mathews, 2000]. The role of public administration in this context will depend on the position of different political forces. It may work to protect people, as well as to throw people in fire. The recent experience of Balkans shows different ways of the involvement of public administration. For example, in Bosnia some local governments collaborated with international NGOs. While in Kosovo foreign organizations were expelled from the territory with the beginning of the conflict. They became active within Kosovo only in the actual post-conflict situation, when the military activities terminated but the inter-ethnic conflicts are still present, and there is the trend for the conflict to "migrate" in FYROM.

Analyzing the role of NGOs in conflict zones, Hulme and Goodhand [2000] have also found that the role of NGOs "*on the dynamics of peace and conflict*" is limited, due to the fact that there are other more important social, economic and political forces active in the conflict: "*NGOs and their activities are only a small part of the overall picture*". Nevertheless, the role of NGOs must not be underestimated, as conclude Hulme and Goodhand pointing out that NGOs strongly linked with other organizations (including governments and donors), and with military groups of both sides, become capable of strategic analyses that lead them to have some influence in political instances. A typical example of involvement in crisis of ICT-oriented NGOs was ZTN ("ZaMir Transnational Net", "za-mir" stays "for peace" in Slavic) during the war in Bosnia. The history of ZTN clarifies many aspects of possible role of NGOs regarding security and collaboration issues. In beginning of 1991 the policy of new nationalist politicians in Ex-Yugoslavia was to cut off all communication means between their subjects, so they could manipulate them easier for future civil conflicts. Monopolized radio-television broadcasting had the same effect in this region. ZTN was created in all the territory of ex-Yugoslavia by a group of hackers as a medium for human rights and antiwar communication. The purpose of the project was to help the anti-war, peace, human rights, NGOs, and media groups in the various countries and regions of former Yugoslavia, and humanitarian aid groups active in the region to be able to communicate better with each other. Additionally, it should help them to communicate with people and groups in the rest of the world ([Bachman 1996], [Skoric 1996], [Bakarsic 1999]). During the hard war times ZTN succeeded with BBS services based on the principle "do not call server - server will call you", to keep for many people an open window of communication with the world. Even after the conflict was closed and Internet introduced into the region, old links of ZTN were used for looking of lost people.

In the middle between "peace and war", there are situated some organizations that operate with "hot topics" for regions with potential political crises. During the Summer 2001 Collegium of SSRC in UC Berkeley, some participants presented the evidence of such "hot activities (extremist activist groups in Ireland, for example - see [O'Dochartaigh 2001]). In this context, the role of NGOs and activist groups may be positive as well as negative. In the latter case, public administrations and governments have their responsibility for balancing the situation and voiding unnecessary clashes.

The role of NGOs in collaboration and security issues, seen in the framework of e-governance, is somehow more than simply an "interface" between public administration and public. NGOs are an important component of the distributed environment making possible the redundancy of network links and of information. This redundancy has two implications. First, contribution to neutralization of manipulation of information that may happen in monopolistic environment. In the Yugoslav scenario, the key in their success in making the war possible was communication or, more precisely, lack of it. Yugoslavia was a country without national television and almost without major national dailies. Once the leadership of different republics turned against each other, they started a vicious propaganda war through the media they controlled. Independent, alternative media were rarely distributed nationally. Major party-controlled media never tried to cross republic lines: they had their target audience precisely defined. The war was then executed out of fear by mostly panicking folks not able to double-check the information they received over government-controlled media [Skoric 1996]. Second, it may serve as backup when public administration IS may go down in crisis time. Such case was the scenario of Albania in 1997, when the public administration services were partially interrupted during the revolt, and missing of this redundancy made the situation much more confuse that it was in reality.

The role of "redundant" NGOs becomes also important because of the growth of information quantity circulating in worldwide networks. Today the main challenge is to select valuable information from available data and processing it in reasonable time. In the information age the data are found everywhere and the added value is growth by selecting and validating these data. In this context the role of information and documentation services changes from simply archiving towards information brokering [NATO 1999]. The role of NGOs would be to serve as "information brokers" for their communities, strengthening the common knowledge necessary for their identity and activism (see [Cwe 2001], for example). To achieve this objective, NGOs need to exchange data between each other, as well as with the public administration. This exchange of data becomes more and more important with the deployment of ICT in DCs.

A "special" security problem emerges with the global use of the Internet - that would be the "public security" on the Internet. The simplest situation relates with the fact that security of one site represents only the half of the problem. When one site is hacked, it may be used as initial platform for attacking other sites. In this context, the security issues get strong public character. But this is only the beginning of the story. Many countries accept that, building ICT infrastructure and facilitating information exchange within the society, they promote the governance and social-economic development. At the same time new technologies have increased the intensity of distribution and communication of socially offensive materials that may be considered as a threat for national identities and cultures. Such phenomena are visible within countries as well as across national borders. Mathews [1997b], for example, argues that deregulation, privatization and globalization helped in transformation of local criminal activities in global enterprises: "g*lobalized crime is a security threat that neither police nor the military - the state's traditional responses - can meet."*

Controlling will require unprecedented cooperation with non-governmental sectors in trans-national scale. Many governments are becoming interested to control and possibly restrict distribution of problematic information through networks within and outside their borders [NRC-CSTB. 1997]. New technologies make more difficult the balance between the control and exploitation, giving priority to the latter. This is a new field of necessary collaboration between governments and NGOs in different levels, as the only way to keep that balance within reasonable limits. The ICT revolution perhaps will favor non-state entities, argues Mathews [1997b]. This is another argument for the necessity of a redundant NGOs "shell" around the public administration systems in the framework of e-governance… On the other hand, concludes Mathews, shifting of power form governments towards other entities in the framework of globalization may lead as well to new conflicts and problems; while increasing voices of individuals and groups with different interests may lead to less common identity and interest for public goods, threatening the democracy itself.

Resuming the conclusions of the analysis, we may say that: first, information systems of public administration have to play a central role in the framework of international collaboration and security. Second, public administration cannot play this role without the collaboration with NGOs and populations, and all these entities with each other in regional and international scale. Third, to address successfully security matters, it is necessary to pay more attention to human factors, remembering that IST must support people and not simply replace them. Fourth, impact of ICT seems to depend not simply in implementation of infrastructure, but also how it is embedded in the society. Former President R. Nixon said: "only people can solve problems created by people" [Wriston, 1997b].

To achieve the collaboration between three entities (government - NGOs - population) and keeping the right balance between democracy, human rights and public security, citizens results the most important but in certain ways "abandoned" entity. It means that not only the technology must match with the existing reality, but the training and education of people is critical as well. "*The objective is to built not only a technical but also an institutional and human infrastructure for data sharing*" [NATO 1999].

5. **IS versus ICT - what is missing in DCs.**

While arguing on e-governance issues in DCs, it would be useful to understand clearly the main factor that makes ICT to have little impact in these countries. It is becoming now a very used question. Many authors do analyze different factors related with this lack of impact, pointing out first of all the missing or inadequate infrastructure and of financial means (see for example [Curry 2001], [Choucri 2000], [Braga 1998] etc.). Of course, infrastructure problems may make difficult the development. But infrastructure problems are result of low general economical development. Impact of ICT goes through transformation of production processes. In normal conditions infrastructure develops gradually in parallel with the general development of the country. ICT are only a necessary condition for the presence of companies in the global market, there are other conditions that make the "sufficiency", as the nature and quality of goods and services produced, banking systems, taxes and customs regulations, cultural and language differences etc. Heeks, Mundy and Salazar [1999] argue that *the key to IS success or failure is the amount of change between "were we are" and "were we want to be"*. In this context, the question may be reformulated in different way - "why there is little development in DCs even when new ICT infrastructure is put in place?" While Choucri [2000] concludes that inequalities are rooted into the global system, and cyberspace related infrastructure is "*positively associated*" with the economic performance.

Taking as model Albania, we will try to reformulate the complex of these factors as seen from inside. The main factor is related with the confusion between the notions of Information Systems and Information technology. Many authors discuss the notion of Information Systems. Heeks [1999a] while arguing on ICT and poverty defines the ICT as "*electronic means of capturing, processing, storing, and communicating information; in order to make this useful, we add in two further components beside the technology and the information: processes of purposeful activity and people to undertake those processes; all of these together now make up an 'information system'*." But, continues Heeks, *this information system cannot sit in a vacuum. It exists within an environment of institutions and of influencing factors.* Hendrick [1994] gives us another similar definition, *considering an information system as nothing more than a set of people, data, and procedures that function together to supply information for decision making… Strictly speaking, an information system is any systematic method of handling data and presenting information.* We need to emphasize the definition of information systems more clearly so we may understand better the impact of ICT in DCs. We will use the logic of "strictly speaking" definition of Hendrick as in Fig.3. Following the idea of Fig.3, we will define the "Information System" as a "virtual engine that collect, store, process data and produce information". IS as a virtual engine is composed by rules, protocols, and procedures based on formalizations. The aim of this definition is to separate IS from the IT and other resources, and to avoid any confusion between IS and ICT.

Based on this redefinition, we may throw a look back and remember the history of the state and public administration in Albania. The state was created in the beginning of 20[th] century, its consolidation started in the years 1930s, and the process was interrupted by the war. After the war the new regime created two parallel structures - state and party - that covered everything. The economy was completely based on predefined plans and enterprises were forced to follow it in any cost. The movements of people and their private activities were strictly limited, private propriety and economic activities were forbidden. In such conditions both in economy and in public administration simple Information Systems were developed, sufficient to record accounting, planning and small movements of people. Almost every 5 years a new wave of "revolution" shook the administration, giving more and more power to the party branches. All current problems were resolved using local improvisations based on ideological criteria.

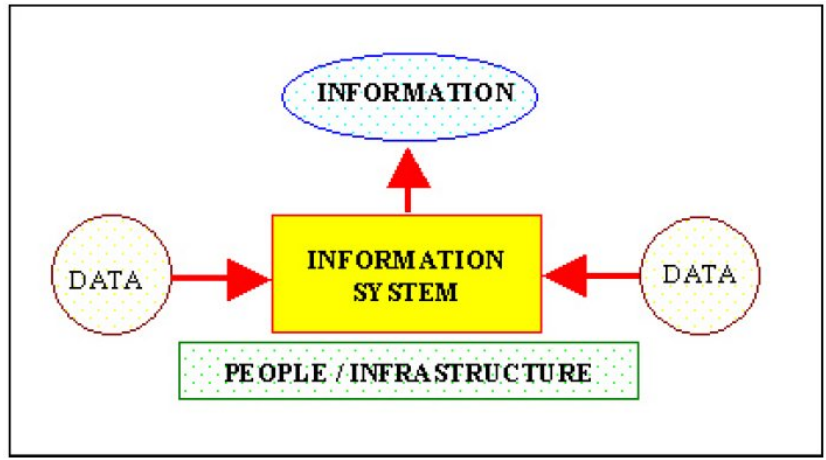

Fig.3 - Definition of "Information System"

Computers were introduced in Albania in 1971 and used for engineering purposes, but not for management - administrations did not have the need of it. Formally it would be used for a better management of country resources, in practice it never happened. During the years 1990s everything from the past was reversed, and administrations work was based partially on tradition, partially on local improvisations controlled by top-managers, and partially on new incomplete legislation. Moreover, many old and experienced people moved from public administration due to different motives or were incapable to adjust themselves with the new realities; while young people have little experience or are educated with western rationality, which does not match well with the local reality [see Avgerou 2000 for alternative rationalities].

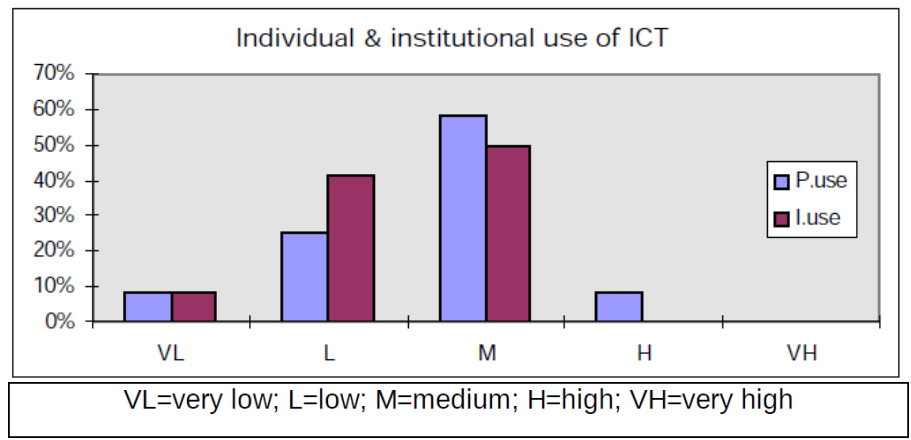

Fig.4 - Usage of ICT in public administration.

As result, Albanian public administration represents a "blurred" environment, where work practices, regulations and procedures are little formalized. If decision-making is based on clearly defined resources, options and tasks, then we have to do with structured decision problems, and the working process is more a routine and can be programmed. Otherwise decision-making is based on ambiguities and we have unstructured decision problems [Hendrick 1994]. Characteristic for Albania is that decision-making has unstructured character, and there is no consolidated and formalized IS in place. In such conditions it is unrealistic to think even about effective institutional automation using new ICT. It may be used to improve and make easier the work of individuals, but in institutional scale the ICT impact is negligible. In Fig.4 we present for individual and institutional applications in Albanian public administration, as seen by in-house IST people.

Individual use of ICT is when people use their desktop PC to keep data and texts, use spreadsheets for some calculations, exchange the files with other colleagues, but all this is organized individually by each person in its own way. Institutional use of ICT means to have built institutionally-unified databases where the main data of the organization are stored, and all the people use intensively those databases for their work, perhaps by mirroring and individually processing pieces of data in their own PCs. But in reality we see the predominance of a wrong feeling that, by using networks for exchange of some .DOC files, it means "institutional use of ICT". It is not a surprise that, when they have to define the objectives of the project, the priority is given always to the infrastructure.

If there is little place for automation, there is less place for 'informatization' of the institution. A number of authors [see Bellamy and Taylor, -1994] consider the public administration as an organization that always has collected, stored and processed quantities of data, and IT has been a useful tool for the automation of information processing. To have effective automation, formalization of structured decision problems is a necessary condition. This is not the case of Albanian public administration, where computers are used, in the majority of cases, for automation of individual working processes, each of them in its own way, without any integrated methodology in institutional scale. Moreover, Bellamy and Taylor consider that "*Informatization occurs*, *when data collected for a multiple of purposes, at different times and places, can be integrated and shared to become resources of vastly increased significance and application*." This is a condition for ICT to have a significant impact in public organizations and facilitate a general leapfrog (i.e. not simply technological). Willcocs [1994] points out that while ICT refers to "*information based technologies*", more important is mapping of ICT into IS composed by "*organizational applications*". There are organizational applications, based on ICT, that by combining and processing organizational data, produce the information necessary for decision-making. In this context, individual applications partially are replaced by organizational applications and partially are integrated in it. Speaking about information revolution, this implies the 'informatization' revolution, i.e. application of ICT on integrated formalization of different individual information processing tasks. Moreover, the process of producing information through integration of applications, a far as expert systems and artificial intelligence are not used considerably, remains relatively not automated and there is the necessity of a quasi-permanent presence of IST expertise within the organization, to achieve or support this integration.

Bellamy and Taylor [1994] insist that it is necessary to distinguish the impact of technology on the government, from the effect of new ways of information processing and communication for the government. Chester [1998] concludes that, while the optimism on ICT issues is leading towards the "*reintroduction of the worst features of the industrial age*", people are confused by the misuse of terminology; while enthusiasts of "information society" emphasis technological capacities for information processing, remaining unaware of the fact that "*improvements in information processing outside the mind does not lead to more meaning within the mind*". This problem is not isolated only in Albania or in DCs. Cases of unsuccessful implementation of ICT is well known even in developed countries (see for example [Margetts 1991]; [Taylor and Williams 1991]; [Willcocs 1994]; Hendrick 1994; Heeks, Mundy and Salazar 1999).

It seems to be a global blurring of these issues by practitioners of ICT. The old and new arguments suggested by many authors seems to be a "voice on desert"… Examples in Albania are typical: A ministry well equipped with PCs and a simple network invests hundreds thousands of dollars to make new network and install big servers, when nobody knows how these servers will be used. Extra installations are motivated with theoretical arguments about the imaginary needs. The network was designed without any consideration about possible flows of data, and the needs of administrations overshadowed those of technological entities. Telecommunication links of 2 Mbps were designed to link remote offices, in a time when those offices exchanged only a few pages of paper per week. The same error was done in the 1980s, when the metropolitan network was designed on the basis of theoretical needs of the administration, and little used by the latter for this purpose. We may justify the past errors due to ideological constraints, but today it is difficult to justify such repeated errors. Nevertheless, this experience was considered as successful and now there is a trend of replication in other ministries.

What is worse, majority of these projects are sponsored and controlled by foreign organizations or governments, following the pragmatic formula "we have a problem, let's get some computers". It is not a justification the fact that objectives of projects were defined through interaction with end-users; and it is not simply to take into consideration user requirements - when unstructured decision problems prevail, even users do not know what they want, or have wrong ideas. Reflecting on such events, a question emerges: "where is the problem, *only* in DCs, or *and* in developed countries?". The warnings of scholars on relations between IT and IS seem to be "materialized" in the Albanian reality, where the feeling is as being within a "gravity hole" - money is spent without being able to "get out". If we combine the logic of Arquette [2001] on "institutional capitals" and of Bimber [2001] on the role of human cognitive phenomena, the chain that leads to the social-economical-political impact if ICT may be represented as follows: Infrastructure => Access => Usage => Cognitive-Phenomena => Impact-on-Work; as in Fig.5.

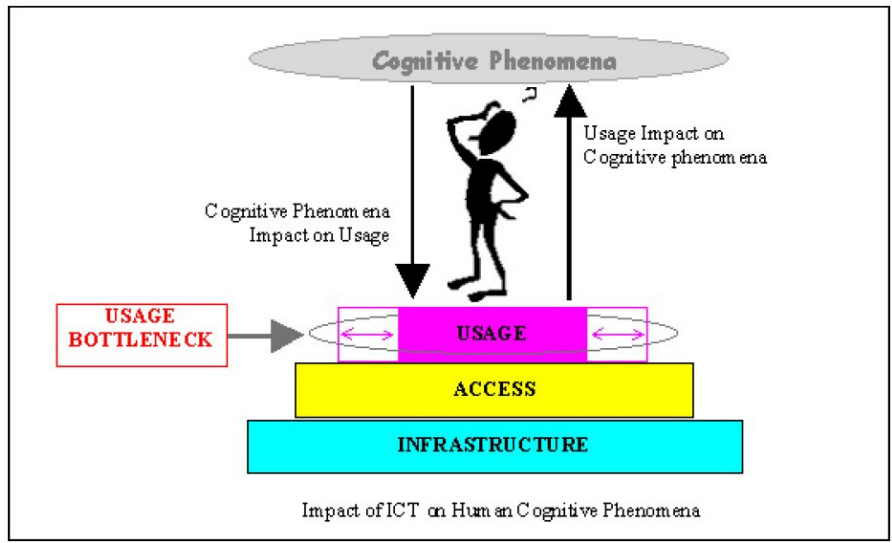

Fig. 5 - Relations between Infrastructure, Usage and Cognitive Phenomena.

The bottleneck in the case of Albania, a well as other DCs, lies in the "middle" of the chain, i.e. "usage". Without a proper usage that can develop such particular cognitive phenomena at the head of people, to lead them in changing for better the way of living and working in community, we cannot have impact of ICT simply by building the infrastructure. This logic leads to three important issues: (1) usage, i.e. use of ICT for structured information systems (IS); (2) education of people using ICT as automated IS infrastructure; and (3) institution building to improve working through deployment of ICT.

The logic implies a solution how to escape from the "gravity hole" related with the confusion of IS and IT, that is "Institution Building". Mainwaring [1998] considers that "*Institutionalization means the process by which a practice or organization becomes well established and widely known, if not universally accepted.*" During recent years many institution-building projects were undertaken to improve and consolidate public institutions. This process addresses both working legislation and working tradition. Legislation had to be modified according new conditions of the country and trends of development, while tradition must break from local improvisations practiced historically. At the same time institution-building has to do with infrastructure issues, and in this context it is important to well balance the quality of the facade of institution with its internal functions. In the framework of such projects a considerable IT infrastructure was installed, and continue to be installed and re-installed, but without being integrated with the institution building process. To build IS in the framework of institution building means to invest in human resources - define how they will work and prepare them to work in the defined way. Moreover, human resources have the potential of to overcome many difficulties [Hirschman, see Salomon et al. 1994]. New presumed effective ICT systems will rely on existing data, systems and processes; but where these elements are not structured, adding ICT may increase confusion and costs. Preparing human resources means accepting the fact that the quality and security of data depend on motivation and values of people involved [Heeks 2001b]. While Schware [2000] points out three key factors of ICT implementation success: for whom (understanding the work environment of end users, needs and benefits), what bundle of services (including training as core activity that enables self-sustaining and links up for ICT applications), and how well projects are managed. All these are more or less neglected by decision-makers, and the priority remains permanently on infrastructure. Facade gets more value than functions; it is a well-known destructive practice of socialist times.

## 6. Strategies and Practices - Good or Bad?

Arguing on importance of social factors, Davison, Vogel and Harris [2000] wrote that it is the social order, with its individuals, groups and institutions, that has the responsibility for the consequences of technological impact in society: "*the notion that technologies can preserve their own course of action is mythical.*" They argue also that the necessary synergy between technology and social context is a non-deterministic one and it requests the adaptation of both technology and social context. Following the same argument, authors at [NRC-CSTB. 1997] point out that the changes, that IT is generating in power relations within governments, between governments, citizens and institutions, impact the power of governments. To achieve national goals, it may be useful if some shift in power occurs, or strategies are modified to match with the impact of IT. The key to success is the anticipation of these changes, for which governments have not been so successful. Technology is non-deterministic and its impact on society will depend on policies and strategies adapted to address these impacts. In this context, crucial is to understand the consequences of the impact in society and particularly in governance, and planning of necessary policy and practice changes.

In Fig.6 we present the level of investments in Albanian public administration, seen from the point of view of in-house IST people. The upper bar chart represents investments in IS, IT and training. The lower bar chart presents several factors identified as cause for bad results in deployment of ICT [Kitiyadisai, 2000]. Lack of significant impact of ICT application in the Albanian reality are result of: (1) misunderstanding the role of ICT and (2) non-changing policies and practices to reflect the new ICT brings in. The first factor is analyzed in previous sections - the confusion between IT and IS. The second factor has to do with strategies and practices of implementation of ICT. The examples of implementation of ICT by the public sector indicate a contradictory slow development through numerous repeated errors and failures.

The typical bad practice is "manipulation of projects". It means that managers who control the program under which the project is situated, use their influence to manipulate with objectives and selection of people or organizations charged with the project. As result one of following may happen: (1) non-experts are involved for the definition and carrying out the project; or (2) experts may be involved but only partially and as individuals; or (3) organizations without expertise or simply motivated by commercialism are involved. As result, the fate of the project is condemned by: (a) badly defined objectives; (b) bad management and realization; (c) expertise involved only on technology issues while the bottleneck is usage; and (d) probabilities for the results of the project to disappear after its termination (nobody takes care of it, nor is able to). We will shortly analyze three typical cases (see also [Frasheri 2000]).

The second bad practice is the consideration of pseudo-experts as experts, i.e. good users of ICT are considered as "good experts" of ICT able to undertake development and deployment of IST. The public administration actually suffers from the lack of ICT experts, because many people have left it or even the countries, as well as young experts try to go abroad for studies and work. In many cases positions of IST staff in public administration are occupied by good users of computers, they have 'product knowledge' but they lack in 'conceptual knowledge' (see [Zehnder 2001] for example), and this handicap does not permit them to see further than a simple use of PCs and MS-Office. It is quite normal they will ask for infrastructure only. Meanwhile the specialized technical institutions are pushed aside.

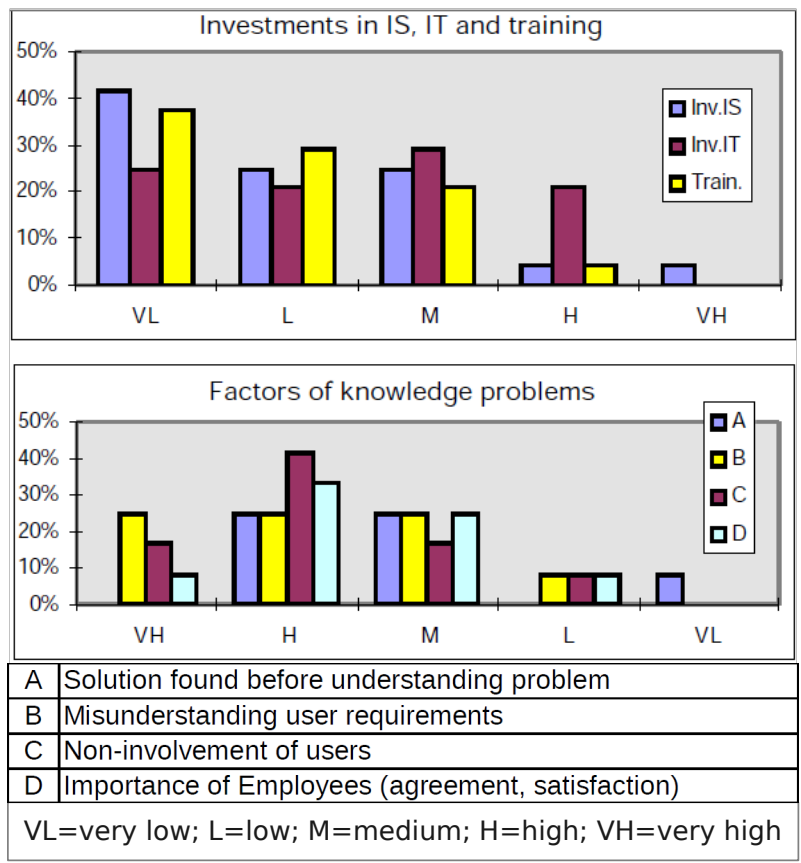

| A | Solution found before understanding problem |
| B | Misunderstanding user requirements |
| C | Non-involvement of users |
| D | Importance of Employees (agreement, satisfaction) |
| VL=very low; L=low; M=medium; H=high; VH=very high | |

Fig.6 - Investments and practices on IST development.

The first case refers to the involvement of Albania into a EU Phare regional project for the development of the academic network in 1995-97. The Phare program was under control of the Committee of Science and Technology. This Committee, under the pretext that the project had to do with "information networks", charged for its realization the Center for Information and Documentation - a small organization under its jurisdiction. The Committee was clearly warned that the project had to do with technological issues and that the center of Information had to capacities to do it; and there were technical institutions of Academy of Science or University that ought to do it. Presumably wishing to have such a "good project" under control, the Committee did not listened to the warnings. The Center of Information showed to be quite incapable to do anything, despite the help from technical community, and the project was doomed.

The second case refers to a big project proposed to one of the ministries by a foreign government. The foreign experts decided with the top managers of the ministry about the detailed technical objectives of the project - that was implementation of infrastructure (high bandwidth links, big servers etc.), without any consultation with local experts and without taking into account the real needs of the ministry. Latter, foreign experts used the strangest ways, bypassing the ministry, to contact local technical people for help. Local technicians evaluated the needs of the ministry and found that the infrastructure proposed was quite unjustifiable, but already it was impossible to modify the project. This time the problem was more serious because in the manipulation of the project participated foreign experts, and this experience seems to evolve in other sectors as well.

The third case is with another ministry. They had an offer from a company to build their internal network and installation of some gig servers, theoretically justifiable. ICT staff of the ministry asked for experts to evaluate the proposal. A working group did an evaluation of different departments of the ministry - working practices and information needs. They discovered that due to working practices, the ministry had no need for any database. The big costly servers were canceled from the project. Some years latter they implemented a good infrastructure given by foreign donors; and hired for the maintenance of its infrastructure a good expert in electronics, with good political links. Sometime after, someone hacked their web-server and changed completely the home page. ICT staff of the ministry had the network out of control and they were unable to do anything, even to put the server off-line. Their ISP complained that the ministry never heard about security matters leaving the server without any protection, and took the initiative to put on-line the mirror of the web pages in its own server.

The problem of project-manipulation and of involvement of pseudo-experts is partially imported in Albania. Pushed by many foreign and international organizations, emerged the practice of PMU - project management units, created within public administration. This lead to practices as described above - pseudo-experts take the control of development and managers have total control of them for manipulations. While in research organizations the new staff is selected carefully for the tasks to perform, this is not the case of public administration where everyone may do every job possible. PMU-s are created and dissolved and from many projects nothing remains in place. The practice of PMU-s may work when the market is consolidated, or there are no other resources. In conditions of transitions, with markets oriented towards selling of equipment as a way for fast accumulation of monetary resources, and when there are other public specialized resources available, a question emerges: may we call such practices as a good sign for development?

It is said that it is necessary to emphasize more on human and organizational issues instead of technology [Heeks 2001b]; and that many times failed IT-based systems have been developed with little consideration on social, organizational and political context [Willcocs, 1994]. It means, first, that the tasks must be distributed according to technical capacities of people (i.e. human issues). Second, to address organizational issues it is necessary to well balance internal in-house expertise, which is more user-oriented, with external expertise technologically oriented. Moreover the external expertise must assure some longevity to assure the continuation of support for the organization. Third, again for organizational issues, it is necessary to take into account the requests of decision-makers; when they do not know what they need (case of bad institutionalization), or they charge lower officials to decide what they (decision-makers) want, the results of presumed "informatization" will be negligible. Some years ago a big Albanian state company asked for help for informatization of some sectors. Managers of the company had no clear idea (from management point of view) what would be done, and they used to change their requirements almost every week. Not only the development of application became an "inferno" for programmers, but both software and documentation final results were considered as non-satisfactory.

Post communist countries had good human resources and R&D institutes able to design and develop qualitatively. New political forces, as well as foreign and international organizations, neglected this reality. Poor in economical point of view does not mean poor in knowledge and technical capability. This logic was totally neglected.  Only recently there are signs that government circles wish to hire local experts for the supervision of the projects. This is only a half step forward, and this is an important issue for the success of ICT implemented. Local experts know better the reality, while foreign experts come with the "western rationality" in their heads (see [Avgerou 2000] for alternate rationalities). Local IST experts may understand better user requirements, what requirements have institutional or individual character, what results from consolidated tradition or well-formulated legislation, and what from local practices and improvisations. Foreign experts in practice are more oriented towards a theoretical evaluation of user needs. In their project drafts, they describe system functionalities that impact the institutionalization; afterwards all the attention points the technology. Institutionalization premises normally are not achieved during the realization of the project, because the latter has to do with informatization and not institutionalization, and as result infrastructure implemented has little impact. This must not be a surprise, Ridge [1994] reminds about the simple fact that the first reason applications do not meet expectations is failure in addressing user needs. While technical obstacles are improbable when up-to-date technologies are used, and this is quite true for DCs. We continue to repeat the errors of the past …

Implementation of huge but not justified infrastructures may have some positive effects that earlier or latter people think about IS. After the first pleasure of playing with big computers, people at the ministry started to think for new projects to develop new IS. Of course this is not the best way, at least economically and for the time it requires. Nevertheless, there are signs of development and of first bricks of the e-governance. We may analyze some good examples. The importance of such examples, even negligible, is that they push concretely towards better governance. The lack of good traditional governance based in a normative framework, phenomena as unstable economy and politics, marginalization of legislation by personal leaderships, all this may lead to very limited e-governance, but it is not a reason for inaction [Heeks 2001b]. Taylor and Williams [1991] observed that while the impact of ICT is low, governments in all levels are more and more considering ICT as increasingly significant, and adopting it as consequence.

The Central Commission for Elections does the recent case of using ICT to communicate with the citizens, regarding the lists of electors. Normally these lists are put in shop windows or building walls. During the past local elections the lists were done using computers and the names sorted by alphabet, anyway there were many errors. While preparing for the new parliamentary elections, the Commission put the lists on its web page, improving the ways citizens and all interested subjects may control their names. The web site has also information on polling rules and procedures, decisions of the Commission etc. It is a good example how a public organization tries to make itself transparent, especially in conditions of a severe political climate.

In other public administration institutions the implementation of IST is going from top to bottom, government institutions having priority compared with city halls. The following data refer June 2001. Ministries represent 40% of domain names registered under "gov.al"; others are government agencies. Within ministries, 80% of them have active web sites, maintained in-house or outsourced to ISP-s. Other 20% have email and/or non-active sites (only ping to such sites receives an echo from ISP). Interesting is that the only ministry which has outsourced all IST maintenance has a domain name but not a web site at all. From city halls, only that of Tirana has registered a domain name, they are working on it but there is no web site yet. Institutions such as of Customs and of Taxes do not have registered any name and they have no web sites. Also social insurances are lagging behind (the social insurance number is not introduced yet, despite several years of working on it). Agencies of Economic Development and of Privatization have domain names and web sites, publishing their activities on projects, tenders etc. Someone else has its web site in construction.

In many public administration organizations there are emerging ideas for the creation of "information centers" equipped with good ICT infrastructure, where citizens may look for information related with the activities of particular institutions. The first steps for creation of such centers are undertaken by foundations as Open Society Institute (Soros), USAID and others in collaboration with public administration of main cities, in parallel with actions for informatization of local public offices. Such projects include the cities of Tirana, Durres, Fier, Berat, Gjirokaster, Kucova. These centers are expected to exchange data with different offices of local administration and facilitate especially marginal communities with necessary information for their relations with public administration. Such information would include data about real estate, police, civil status, social status, finance, taxes etc.

The future of these centers is not clear. There are two main problems identified - sustainability and legitimacy. Actually these centers are sponsored by NGOs, and their activity supported by local authorities. The sponsorship from NGOs may not long forever, and self-financing may be a problem due to the fact that these centers are dedicated mainly for poor people. The legitimacy of their activities has to do with the fact that they will remain simple centers of information where people may learn about their status and possible solutions, or their information will have a legal value so people may conclude their affairs there without being necessary to run from one office to the other. It remains the time to show how these centers will progress.

In the framework of e-governance, the role of NGOs does not terminate with the sponsorship of some projects. As Madon [2000] argues that because NGOs may operate simultaneously at different levels of governance, they have the capacity to link low-level experience with high-level politics. NGOs may serve as interface between government and different communities. A special role of NGOs (as well as private service providers) in this context is creation of networked centers to help a widespread community of poor people, whose lacking of money to acquire ICT may become obstacle for extensive deployment of new technologies [Heeks, 1999a]. From the Albanian reality we have a good example of the Open Internet Center of Soros Foundation, which created conditions for many students, teachers and researchers to have Internet access during the years 1997-99, and still supporting with connectivity a part of the Academic Network. The creation of private ISP-s in Tirana was followed by many "Internet-Cafés" becoming very popular, especially between the youth.

All these good examples cannot give the answer for the question of ICT lack of impact in DCs. While DCs may leap jumping "highly" but little in advancing, developed countries are crawling too quickly. There are two possibilities for DCs - or they will improve their "leaps", or lets hope that in developed countries there would be (if) some slowdown of their rush towards the "promised" information society. At least we may calm down our anguish referring to the Riepl's law of complementary, that during the human history none of communication modes has disappeared completely, but only was adapted to the new realities and complemented by new modes [Kaase, 2000]. Anyway, one thing is certain - DCs need to improve their leaps, and this lead us to argue about the strategies.

The question has to do with the relations between top-down and bottom-up ways of development. We saw that the premises are for decentralization of public administration, that pseudo-experts often occupy key positions of ICT staff, and that big projects are likely to be manipulated. These premises lead us to think in favor of bottom-up development. "Mathematically" that means to do local optimizations, and this may be fatal for the global optimization. So we need to match together both bottom-up and top-down ways of development. This argument is for a "central coordination" in a "distributed environment" we have talked in previous sections.

The solution would be a "national policy" for the development of IS over modern ICTs in public administration.

The feeling of in-house ICT people in Albanian public administration on these issues is presented in Fig.7. The upper bar chart represents the weight of some factors that negatively impact the deployment of ICT [Kitiyadisai, 2000]. The lower bar chart represents the feeling about top-down and bottom-up development strategies [Dlodlo, Ndlovu, 2000].

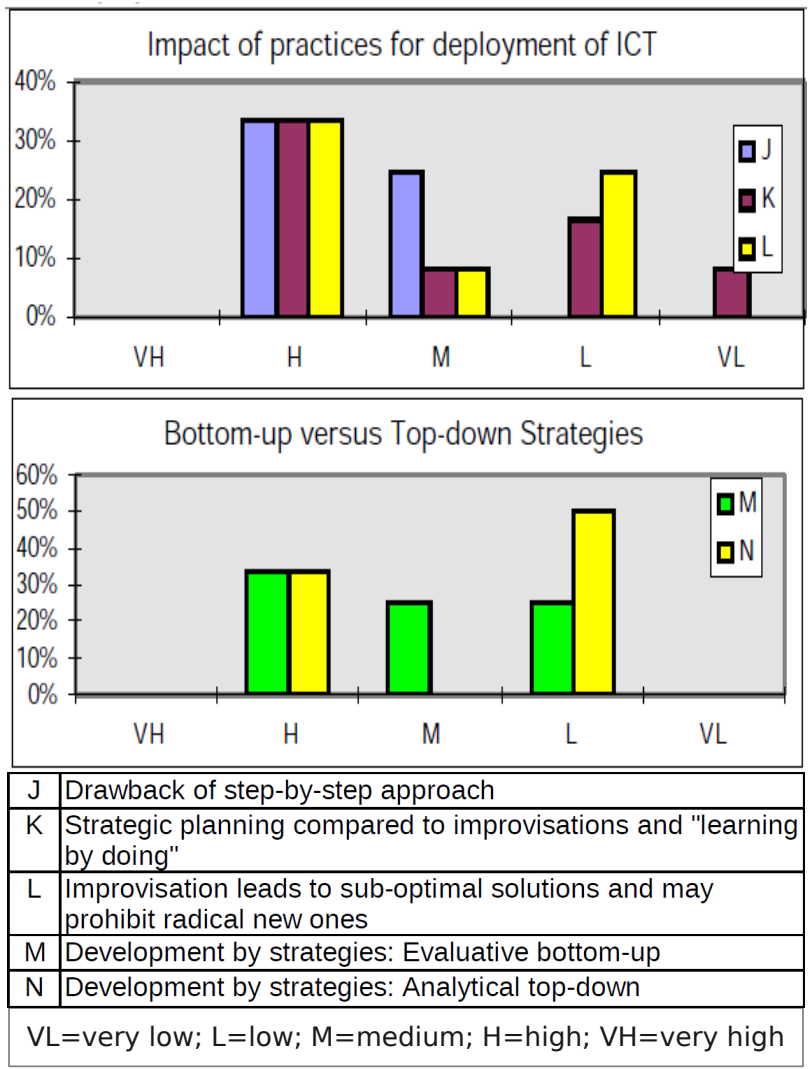

| J | Drawback of step-by-step approach |
|---|---|
| K | Strategic planning compared to improvisations and "learning by doing" |
| L | Improvisation leads to sub-optimal solutions and may prohibit radical new ones |
| M | Development by strategies: Evaluative bottom-up |
| N | Development by strategies: Analytical top-down |

VL=very low; L=low; M=medium; H=high; VH=very high

Fig.7: Practices and strategies on deployment of IST

From the charts is seen that the attitude of IST people is contradictory. In one hand, they tent to evaluate negatively different aspects of bottom-up strategies, while under-estimating top-down strategies.

Without under-evaluation of bottom-up initiatives as crucial to address the growth demand for interactive services in an emerging networked society [Fair 1996], we need to emphasize the importance of some development strategies. Many authors discuss issues of strategies and policies. Bellamy and Taylor [1994] for example, remark that it is the "information policy" missing in the framework of research to give the information perspective. Following arguments of Hashem [1995], Davison, Vogel and Harris [2000] argue on creation of a regulatory environment, based on analysis of technologies before their application, to avoid uncoordinated decisions that may lead to costly and useless solutions.

 Mitra [2000] analyses examples of ICT development strategies in India, pointing out the importance of strategies to deal with the fast increasing ICT sector and compete for more investments. Moreover, Mitra mentions the case of Andhra Pradesh where IT state policies address with both attraction of private investments and development of public sector ICT policies and projects. Following with key components of IT policies, Mitra suggests: promotion of private investments though promotional activities and schemes, investment in physical infrastructure (sectors as real estate, power and water, transport, telecommunication), and investments in human resources (research, education and training), development of IT applications and content in the government sector.

All this is a complex development touching many sectors. The government must collaborate with public organizations, in particular with R&D community, to prepare such strategies that serve the development of the society. The idea of creation of government agencies for this purpose - development and application of strategies may be a risky one, becoming an obstacle for the decentralized development because of corruption and manipulation of projects. Important is the exchange of information between independent organizations, following a certain strategy that puts some common guidelines, aiming to assure the interoperability of separate "bricks" of IS and ICT to assure their function as a single body. The Internet itself is a good example of independent networks interconnected using the same protocols of communication. This idea must be extended not only for the infrastructure but also for IS - independent systems following compatible rules, protocols and procedures to assure their interoperability.

It is necessary to well balance both top-down and bottom-up ways of development [Hendrick, 1994]. The idea as seen in the Albanian reality is discussed in [Frasheri 2000]. Heeks [1999b] also makes a detailed analysis of centralized versus decentralized approaches, pointing out that, in order to overcome "challenging contradictions" seen in the application of centralized and decentralized approaches, one way would be a "core-periphery approach that attempts to reconcile the push of the centralized approach with the pull of the decentralized approach." This combined approach, continues Heeks, may be achieved through integration of both approaches, as well as through division of their spaces of application. The same argument of balancing core-periphery / centralized-decentralized / top-down-bottom-up approaches seems to be valid even for developed countries in their way of globalization and strengthening of democracy [Cederman and Krause, 2001].

The approach is not an easy one due to the complexity of problems in social and political aspects. Technical principles may be simple - we need good internal databases in public administration institutions, and good web-like interfaces to access non-confidential data. Important is the structure, classification and indexing of data, so they may be connected with each other. The same is valuable for NGOs and enterprises. These distributed databases may communicate with each other; each organization may acquire exact and complete information about their working environment, and make possible the coordinated development in decentralized environment - that is one aspect of e-governance. In this context, the idea of forcing a strategy does not mean control from the top. Instead, as Landsbergen and Wolken [2001] argue, it consist in "recognizing that in a technological society technical standards are another kind of public law", i.e. defining and applying without exception some principles that will lead to: (1) improving the functional requirements of organizations; (2) building infrastructure depending on the real needs; and (3) assuring the interoperability of systems and networks.

## 7.  Public or Private - that is a question.

 Many authors discuss the relations between public and private sectors with implications on implementing ICT in public administration. Reflecting the Albanian reality, in [Frasheri 1998] we suggested the necessity of a balance between these two sectors. Pfaff [2001] writes that even in developed countries as UK with a consolidated private sector, services for citizens have suffered because priority was given to private sector in name of eliminating bureaucracies and reducing taxes. The Albanian reality proves that existing trends lead towards a polarization of the society in a dipole Administration <=> Private without space for other structures as public academic institutions. There are two questions related with these trends: (1) how much the ICT development is outsourced to private firms, and (2) how public administration collaborates with public academic institutions. The question of academic community for small countries as Albania may seem irrelevant, but it is related with the quality of high education, and indirectly with the education in general. Moreover, in certain realities as in Albania, academic public institutions have the budget from the state, but administratively they are autonomous organizations; they are positioned somewhere between the state and NGOs, but unfortunately treated as "pure" state organizations. Penalizing directly or indirectly the education, we create a fragile background for future development, and good probabilities to remain forever in a "gravity hole" as we are today. Someone may call cyber-space as "the great leverer", it simply lever some social differences to replace them with something else - knowledge-based differences. "Investment in education is the fundamental source of national power" [Choucri, 2000]. Even today many people have wrong ideas what the Internet they use daily is.

Balancing in-house, academic and private experiences is also a necessity for a sustainable development, increasing the independence of public administration from market oscillations. In the Albanian public administration the IST maintenance costs are much lower that when private firms are contracted. A simple solution is to create legal conditions for public administration to collaborate directly with public R&D institutions without tendering for certain categories of projects. Another argument is linked with the essence of "private" itself. The main cause because private activities develop better than public ones is the economical feedback - in private activities this feedback is a direct one - punishes or encourages without mercy; while in public activities it is indirect, moreover its effects may be reversed through political links (typical phenomena for socialist-like regimes). The public sector functioning may be improved if mechanisms of direct economical feedback would be implemented. In post-socialist countries there are examples of such mechanisms used in academic institutions.

Experience shows that the best opportunities to achieve a balance between techno-economic developments are the initiatives between public and private sectors [Fair 1996]. Beside these initiatives, it is important to define ICT policies avoiding the exclusivity of private sector and allocating a considerable part of public funds for facilitating ICT use and for massive training on ICT skills [Hamelink, 1999]. Hill [1996] remarks that sometime cost structures of private agencies work against sustainability, and in such cases even the argument of low cost fails. Willcocs [1994] has already warned the risk of giving priority to private sector for informatization of public sector. Years earlier, while evaluation the results of IT-s in public services, Margetts [1991] warned for consequences of delegating design and specification of IS to external consultants while relatively excluding internal specialists.

Despite early warnings and increased evidence on the necessity of a balance between public and private sectors, foreign agencies working in DCs follow a neo-liberal agenda oriented towards replacement of public sector with the private one [Heeks 1998a]. But these sectors have different complexities [Hendrick 1994], and we cannot replace a complex system with a simpler one without creating new problems that may be fatal in future. Hendrick points out also "*information systems are much more than computers and software*." In this context remains also Willcocs [1994], arguing the necessity of keeping in-house some critical capacities related with IST in public administration. Willcocs further suggest parallel running of both outsourcing and in-house development. Brown and Brudney [1998] evaluate the postulate "*government should seek market rather than administrative solutions to facilitate the delivery of services*" and conclude that: "*while outsourcing is thought to be cost-efficient, facilitate quality of services, allows managers to concentrate on organizational activities, it still must be monitored*." In this context they points out arguments of researchers as Ferris and Graddy that lack of involvement of in-house specialists in implementation of ICT projects may have serious consequences on its results.

Brown and Brudney [1998] emphasize the fact that many researchers consider as crucial in-home capacities on organization, project management, strategic planning, and team work; researchers conclude that the primary reason of IS failures is poor project management. Contracting (outsourcing) of development is supposed save costs and free institutions from investing internally in IST development, while yielding operative systems that better meet the requirements. Research suggests that contracting may undermine management capacities resulting in loss of potential benefits of systems. Brown and Brudney conclude that modest contracting may yield good results for government, but this goodness may decline when increasing outsourcing due to complications of implementation processes. High levels of outsourcing not only undermine positive results of projects, but also block development of in-home capacities for project management and supervision. Moreover, IST projects need to be supervised from administrative and technical point of view, and the latter is decisive for the results of the project. Brown and Brudney conclude that many researches point out the fact that institutions practicing high levels of outsourcing may neglect development of internal capacities important for positive results of ICT deployment.

The necessity of keeping in-home some critical ICT management is also related with continuity issues. Ridge [1994] concludes: "*One lesson is that applications can never be regarded as fully developed and must be continually evolved.*" Moreover, Ridge points out that in national scale there would be a number of separate specific applications for specific users and it is necessary a coordinating framework to assure the interoperability of these applications. This issue has to do with the balance between immediate interests and future interests. If we think of a good government, it also means governance of the present for a future. The "socialist experience" of post communist countries is an example how bad may be the future when governments think only partially (i.e. ideologically or technically) for the future, neglecting social issues. These countries were "crashed" because in time a critical gap was created between politics and society.

Another important issue is the security of IS implemented in public administration. This issue becomes more critical with development of public institutional applications and databases and their interfacing with the world outside public sector. The transitive reality in many DCs, including Albania, is characterized by: (1) the IST private market is not yet well developed, and (2) the justice system is relatively weak and problematic. In such conditions, an opened question emerges: may the risks for unauthorized access to public networks and data increase when development and maintenance is outsourced to private companies (local or foreign ones)?

In Albania we find both practices, i.e. in-home working groups and outsourcing. Many of public administration institutions have their minimal ICT staff. In most of the cases only people with production knowledge compose this staff. As described earlier, this is one of the reasons why projects go technology-oriented. In-home ICT staff does not have the necessary vision to see farther and wider. It is better than nothing - in some cases ICT people feels the need for expert help and asks it. There are also cases when all ICT management is outsourced. There is nobody within the organization who knows what is going on there on ICT matters. This schema may work when ICT is used individually, and when maintenance attention is limited in keeping the network, PCs and MS-Office running. Organizations using market-oriented outsourcing were involved in preparation election lists, which resulted with many errors that became a political problem. Total outsourcing may be very dangerous for the academic community as well. Instead of keeping public funds within the community for its development, money flows to private companies while academic brain evaporates from the country. This penalizes especially universities, with grave long-term consequences on the national education system.

Undermining universities may be fatal for the future of a country. It relates with the whole education system of the country, especially in ICT matters. Willcocs [1994], for example, emphasizes the importance of developing capacities of managers of public sector in formulating of IT strategies as a combination of institutional objectives, as well as testing capacities to evaluate in practice the results of projects. Quadri [2001] points out that while middle management need some more knowledge on ICT matters, the biggest deficit is with senior managers who need to know how ICT may help do advance their business. And there are not only the managers - "*the information society must become a 'lifelong learning society', which means that the sources of education and training must be extended beyond traditional institutions to include the home, the community, companies and other organizations*" [Forum 1996]. The importance of knowledge for the societies of today and tomorrow is crucial for sustaining economic growth and welfare in a context of globalization, as argues Dieter [2001] considering that if national governments would manage to build robust knowledge-based societies, globalization may become an opportunity rather than a threat.

Such objectives cannot be achieved when developing trends of the country lead to a university department of computer science so under-staffed that and someone even thinks to close it as the only solution. It would be a "killing solution", not a "saving solution". Education is decisive in deployment of ICTs. Castells [1999] remembers that like a sword with two edges, ICT creates possibilities of DCs to make leapfrogging economical development and compete in the global market, at the same time for economies unable to adapt with new technologies the slow-down accumulates increasing the digital divide. Furthermore, the ability of the whole society to adapt with the information revolution depend on its education how to "*assimilate and process complex information*". This implies the whole educational system of the country from the bottom to the top, and it is not a question of some elected managers. The necessity for a global and effective education must be reflected in the national strategy for ICT development.  Preparation of such strategy results as a political problem, rooted deeply in political mediums.

## 8.    Conclusions

We agree that many DCs are making progressive steps. But the question of gaps between them and developed countries remains open. We invest in ICT infrastructure without taking care how and why it will be used. We give priority to private sector, and as result one of marginalized communities is just the academic community, in whose hands rests at least the education of society. For DCs to go forward, it cannot be without learning people, especially in ICT. The Internet itself, for example, developed in a bottom-up way, beginning in laboratories and extending in economy and governance, and the cycle closes by returning to laboratories for new developments. In DCs Internet was "imported", in some ways it was introduced in laboratories, extended towards the economy and governance, but here the cycle seems to be interrupted - there is no relevant returning to laboratories - the initiators remain outside.

Subbaih [1999] has its rights to be pessimist. In 1993-95 we worked hard to get a 64 kbps link with the world, today there are projects as Internet-2 in US and Geant in EU, where they use hundreds of Mbps. Based on such links, web sites are developing complex pages that cannot be downloaded with a link of 64 kbps shared between many users. Being not able to follow the running pace of developed countries, the gap is widening. Decision makers in DCs may agree to send 2 Mbps at the office of a senior manager, but they may neglect to do it for a laboratory.

Developed countries and international organizations are helping universities in DCs, and promoting the return of young specialists in their countries, but at the same time the global tendency of developed regions is to attract talents from all around the world, especially in ICT fields [Castells 1999]. As result of this phenomenon, Albanian "brains" are evaporating, especially youngsters from the ICT sector. Someone is even pleased with the creation of a database with names of learned people emigrated abroad. To make the situation more spicy, some emigrant returns in his country of origin, in the same office where had worked before, but now as consultant of international organizations, creating negative impact on motivation of local people. That is to build with one hand and destroy with the other. Again we need a global political to promote with priority the human resources in DCs.

Evaluating the comments of many researchers of social sciences during last ten years, and also the actions of producers (from developed countries), IT experts and decision makers, it seems that emerges the evidence of another gap in human society. It is a gap between social sciences, technology makers and decision makers in developed countries. We have the feeling that in practice the warning of social science are neglected or misused when actions to help developing countries are prepared and executed. The World Bank, for example, is investing a lot in developing countries, but its work practices based on principle of "go private" not always are compatible with local realities. European Union has its requests for countries applying for membership, and in its "own way" Albania is trying to fulfill them, but we do not see any request regarding the IS in public administration (investments mainly in infrastructure continue without looking at usage, training is not always sufficient, education in all levels is becoming problematic), or real improvement and motivation for academic institutions. Something is missing in the chain of links between developed and developing countries, unfortunately a warned missing. The gap between social sciences, decision-making and technology implementations is reflected in the story of technology assessment movement. The Office for Technology Assessment, created in 1974 as an advisory entity for the American Congress, was not able to resist new policy-making ways and ceased its activities in 1995, just in the time when Internet entered in the market [Bimber 1998, Hill 1996]. A proverb says, "One is incorrigible if falls two times in the same hole". It seems that the "incorrigible" element is within the human society.

The questions are how to blend together the decentralized bottom-up trends of development with the necessity of interoperability of systems and networks, i.e. achieving a "centralized" coordination in a decentralized environment; how to make a reasonable horizontal decentralization by well-balancing vertical control and monitoring; how to neutralize trends for polarization of power, negligence and manipulation. The answer to such questions seems difficult, as "to tie a bell in a cat's tail" …

It not simply ICT we need for the development of the country. First of all we need the political will to develop and integrate the country, and clear political objectives how to do it. Second, we need a development strategy of IS as part of institution building strategy, defined on the basis of those political objectives. Third, we need close collaboration with both private and public ICT specialized sectors, both for implementation of the IS strategy and for massive training in all levels. The deployment and impact of ICT will come as result of these combined activities.

*Acknowledgement: This research was assisted by an award from the Social Science Research Council's Program on Information Technology, International Cooperation and Global Security – Information Technology, International Cooperation and Global Security Program Fellowship 2001.*

**Remarks:**

1. EJISDC - Electronic Journal on Information Systems in Developing Countries.
   A joint publication by: University Malaysia Sarawak, City University of Hong Kong, Erasmus University of Rotterdam, the Netherlands, Delft University of Technology, the Netherlands.
   http://www.is.cityu.edu.hk/ejisdc/ejisdc.htm

2. IDPM - Institute for Development Policy and Management.
   The University of Manchester, UK.
   http://idpm.man.ac.uk/idpm_dp.htm

3. BRIE - Berkeley Roundtable on the International Economy.
   University of California, Berkeley.
   http://brie.berkeley.edu/~briewww/